\documentclass[letterpaper,journal]{IEEEtran}

\usepackage{amsmath,amsfonts}
\usepackage{algorithmic}
\usepackage{array}
\usepackage[caption=false,font=normalsize,labelfont=sf,textfont=sf]{subfig}
\usepackage{textcomp}
\usepackage{stfloats}
\usepackage{url}
\usepackage{verbatim}
\usepackage{graphicx}
\usepackage{hyperref}
\def\BibTeX{{\rm B\kern-.05em{\sc i\kern-.025em b}\kern-.08em
    T\kern-.1667em\lower.7ex\hbox{E}\kern-.125emX}}
\usepackage{balance}
\usepackage{tabu}                      
\usepackage{booktabs}                  
\usepackage{lipsum}                    
\usepackage{mwe}                       
\usepackage{ccicons}                   
\usepackage{soul}
\usepackage[normalem]{ulem}
\usepackage[table]{xcolor}
\usepackage{fontawesome5}
\usepackage[most]{tcolorbox}
\tcbuselibrary{breakable}
\usepackage{enumitem}
\usepackage{mathptmx}                  

\graphicspath{{figs/}{figures/}{pictures/}{images/}{./}} 
\definecolor{audience}{HTML}{e0c813}
\definecolor{setting}{HTML}{6cc1e0}
\definecolor{impression}{HTML}{9f76c3}
\definecolor{lightgray}{HTML}{F3F3F3}
\definecolor{lightyellow}{HTML}{FEFFE8}
\definecolor{periwinkle}{RGB}{204,204,255}

\definecolor{periwinkle}{RGB}{196,200,255}
\sethlcolor{periwinkle}

\sodef\guideso{}{0.04em}{0.25em}{0.35em}

\newcommand{\guideboxtitle}[1]{%
  {\sffamily\bfseries\fontsize{7.5pt}{8.5pt}\selectfont
  \scalebox{1}[1]{\guideso{#1}}}\par%
}

\newtcolorbox{guidebox}{
  breakable,
  enhanced,
  colback=gray!5,
  colframe=gray!45,
  boxrule=0.5pt,
  arc=2pt,
  left=6pt,
  right=6pt,
  top=6pt,
  bottom=6pt,
  before skip=6pt,
  after skip=6pt
}

\DeclareRobustCommand{\iconaudience}{%
  \begingroup\normalfont%
   \hspace{1pt}\raisebox{-2pt}{\includegraphics[height=1.5\fontcharht\font`\B]{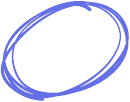}}\hspace{0pt}
  \endgroup
}

\DeclareRobustCommand{\iconsetting}{%
  \begingroup\normalfont%
   \hspace{1pt}\raisebox{-2pt}{\includegraphics[height=1.5\fontcharht\font`\B]{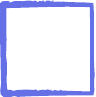}}\hspace{0pt}
  \endgroup
}

\DeclareRobustCommand{\iconimpression}{%
  \begingroup\normalfont%
   \hspace{1pt}\raisebox{-2pt}{\includegraphics[height=1.5\fontcharht\font`\B]{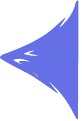}}\hspace{0pt}
  \endgroup
}

\DeclareRobustCommand{\iconaspiration}{%
  \begingroup\normalfont%
   \hspace{1pt}\raisebox{-2pt}{\includegraphics[height=2\fontcharht\font`\B]{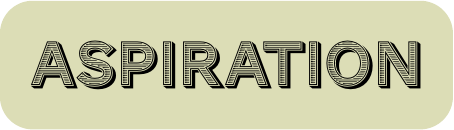}}\hspace{0pt}
  \endgroup
}

\DeclareRobustCommand{\iconreality}{%
  \begingroup\normalfont%
   \hspace{1pt}\raisebox{-2pt}{\includegraphics[height=2\fontcharht\font`\B]{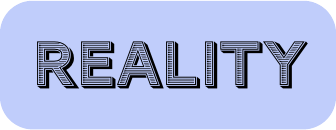}}\hspace{0pt}
  \endgroup
}

\begin{document}

\title{Projecting the Public: Reflections and Recommendations for Empirical Visualization Studies of General Audiences}

\author{%
  Mandi Cai, Jack McNiel, Erik C. Nisbet, and Matthew Kay
  \thanks{All authors are affiliated with Northwestern University.}
  \thanks{Manuscript received xxx; revised xxx.}
}

\maketitle

\begin{abstract}
    Calls for shifting \textbf{visualization research to general audiences} outside of professional contexts have been put forward since the early 2000s, but expanding outwards means contending with a diversity in visualization's communicative purpose beyond transmitting insights from data. After our own experiences studying public-facing visualizations, we sought a set of considerations to help early-career visualization researchers reason through designing empirical visualization studies for the general public that \textbf{confronts the various ways visualizations are interpreted}. We examined literature on guiding paradigms of research, reviewed 73 visualization papers that specify a focus on audiences broader than domain experts, and conducted 15 semi-structured interviews with authors of such papers, to derive the following: (1) \textbf{a conceptualization of any experiment as 3 elements}---casting audience, constructing experimental setting, and projecting audience impressions and interactions into measurable constructs 
(2) an account of interviewees' aspirations for each element of study design, versus the real challenges that occurred. We detail their approaches to \textbf{navigate the gaps between aspirations and reality} when working with general audiences and \textbf{summarize them as actionable recommendations}. The \textbf{conceptual framework} is intended to help junior researchers \textit{scope} and \textit{outline}  experimental designs of visualizations ``for the masses'' \cite{pousman_casual_2007}, while the \textbf{recommendations} provide concrete guidance for putting those designs into practice. The 73 papers are filterable and visualized on our \href{https://mucollective.github.io/projecting-the-public/}{study website}, and our code to sample the papers is on \href{https://github.com/MUCollective/projecting-the-public}{GitHub}.


\end{abstract}

\begin{IEEEkeywords}
    Visualization research methods, study design, empirical studies, general audiences, public-facing visualizations, interview study
\end{IEEEkeywords}

\section{Introduction}

\begin{quote}
    \textit{Turning something from an idea
    into a reality
    can make it seem smaller.
    It changes from unearthly to earthly.}
    
    \textit{The imagination has no limits.
    The physical world does.
    The work exists in both.}
    
    ---Rick Rubin, The Creative Act: A Way of Being \cite{rubin_2023}
\end{quote}

With the start of the Information Age in the mid-20th century and the rise of the Internet in the early 2000s, information visualization and interactive visual tools began their trajectory towards an optimistic vision of mainstream adoption \cite{hutchison_creation_2008, danziger_information_2008} outside of professional and highly technical contexts. Since then, there have been numerous calls within the visualization research community to consider how the general public can and does use visualization in everyday contexts \cite{fu_more_2024, lee_reaching_2020, pousman_casual_2007, danziger_information_2008, viscomm_2025}. Conducting studies of public-facing visualizations---for vast, diverse audiences that encounter visualizations in fluid public forums---pushes the visualization research field to consider \textbf{new lenses through which to understand the ``success'' of a visualization}, \textbf{beyond the effective transmission of particular data insights.}


In our own experiences studying U.S. elections visualizations \cite{yang_swaying_2023, cai_2026}, we faced challenges that seemed especially prominent to working with broader audiences in real-world, public settings---like picking a representative audience to focus on when our display was intended for anyone interested in following a national event (versus, for example, a bounded group of data scientists at a company); dealing with varying data and technical literacy levels, usage patterns, and motivations \cite{lee_reaching_2020} from viewers; designing testable visualizations that conformed to existing visual standards of legitimacy and audience expectations; and picking measurable dependent variables when our outcomes of interest extended beyond successful completion of a data analysis task. Given these practical struggles, we sought to catalog experiences from other researchers to illuminate important considerations for planning empirical visualization studies\footnote{In this paper, we consider study designs that involve presenting a visualization to members of the broader public, along with either an empirical evaluation of the visualization or data collection supported by it.} of broader audiences, with the goal of supporting early-career researchers in structuring their experimental designs. 

Casting a wide net, we scoured existing visualization research literature in the last 5 years that \textbf{did not specify designing for domain or data experts in the motivation}, to collect examples of experimental design and compare and contrast different approaches. The resulting \textbf{73 papers} spanned simulated lab studies to in-situ deployments in the field, in settings like museums, news media, public parks, classrooms, and online web browsers, with participant measures like trust, political attitude, and (of course) performance on analytical tasks. This research review, in tandem with \textbf{15 semi-structured interviews} with visualization researchers who had authored such studies, gave us a grounded look at the behind-the-scenes processes that researchers have actually employed to design and manage these studies.

Derived from our interviews and close reading of the 73 papers and experimental design principles, we present an illustrated 3-part conceptualization (Sec. \ref{sec:framework})---\textbf{casting audience}, \textbf{constructing experimental setting}, and \textbf{projecting audience impressions and interactions}---as a tentative but enthusiastic attempt at a metaphorical framing to support researchers in designing studies for the general public. Researchers engage in acts of \textit{scope}, \textit{disruption}, and \textit{simplification}: casting an audience for the visualization that is representative of a larger population, constructing an experimental setting to study participants in, and projecting audience impressions and interactions into participant measures. 
The conceptualization should be used at the start of study design and when components change (for example, due to practical constraints) to check whether they still \textit{align} with each other.


In Sec. \ref{sec:exp-vs-real}, we illustrate tensions of \textbf{aspirations versus reality} that our 15 interviewees underwent, organized by the 3 elements in the conceptualization (Sec. \ref{sec:framework}). We detail how our interviewees hold aspirations for visualization studies with the general public, like capturing enough diversity across attributes in the participant sample or experimenting in public spaces that general audiences pass through organically. When pressed with practical limitations, our interviewees employed particular strategies and rationales in light of these tensions to mold a finished study design they could still execute. We summarize them as \textbf{actionable recommendations} to support junior scholars in formulating  their own study designs.


Our ending discussion describes interviewees' perspectives on how the visualization community supports the mindset and methods that this line of research demands---provoking a call for action to embrace different kinds of visualization research to accommodate the general public.

\section{Related Work}

\subsection{Visualization for the general public}
\label{sec:related-work-broader}

\textit{How has visualization for the general public been conceptualized, and what notable challenges are introduced?} Considering visualizations for general audiences seems to have stemmed from a recognition that mainstream adoption of infovis, for both personal and professional reasons, could occur in the early 2000s \cite{hutchison_creation_2008, danziger_information_2008}. In 2007, Pousman et al. \cite{pousman_casual_2007} proposed a subfield of infovis research---termed ``Casual Infovis''---to account for user populations of all levels of expertise in both professional and everyday situations, who might use visualizations \textbf{for fun or out of curiosity rather than for professional tasks}. The idea of a general audience as having \textbf{varying levels of data and visualization expertise} is echoed by Heer et al. \cite{hutchison_creation_2008}, Lee et al. \cite{lee_reaching_2020}, and Schuster et al. \cite{schuster_who_2023}. Studying the general public could, in practice, mean reaching audiences \textbf{through mediums for the masses like news media and museum exhibits} \cite{fu_more_2024, dasu_sea_2021, ma_decoding_2019}, or reaching \textbf{historically underrepresented audiences in visualization research} \cite{peck_data_2019, arunkumar_lost_2025}.

Expanding the imagined audience requires grappling with different motivations and settings in which people encounter a visualization---individuals may exhibit \textbf{alternative modes of engagement} \cite{lan_kineticharts_2022, wang_emotional_2019, morais_exploring_2022, liem_structure_2020, danziger_information_2008, schuster_who_2023}, not necessarily seeking to rigorously perform data analysis with the visualization to derive insights. Various taxonomies have been proposed to categorize low-level and high-level goals and tasks when individuals view and interact with visualizations \cite{amar_low-level_2005, sarikaya_scatterplots_2018, lee_task_2006, burns_how_2020}, but many such taxonomies are more applicable to data analysis workflows \cite{amar_low-level_2005, sarikaya_scatterplots_2018, lee_task_2006}. Usage patterns with visualizations in everyday contexts could be momentary (from a TV ad), intermittent (checking an online news dashboard), on-the-go (through mobile phone applications), or over a longer period of time (through repeated visits to a museum) \cite{pousman_casual_2007, lee_reaching_2020}, and deploying a public-facing visualization might involve spontaneous visitors of disparate backgrounds who attribute different personal meanings to the same visualization \cite{sprague_exploring_2012}. These serendipitous encounters, and ``distractions'' in everyday contexts, can complicate (or enrich) a study plan. 

To be clear, we do not necessarily think that visualization researchers should rely on the word ``general audience'' to anchor their study’s target population. In practice, no singular study can solicit responses from every representative group in the general public for evaluation, and the notion of universal design heuristics is an unrealistic goal \cite{burns_who_2023, akbaba_entanglements_2025}. However, we do believe there is an orientation to visualization study design that is meaningfully different when considering public settings and broader audiences outside of data analysts and technical domain experts---it means re-imagining existing academic and professional bounds of what visualization even is \cite{pousman_casual_2007}, meeting people in non-work and mediated communication settings, and considering how to attract participants who are not bought into visualization use by default. We expand on this more in our interview findings in Sec. \ref{sec:exp-vs-real}.

\subsection{Guiding research paradigms}
\label{sec:related-work-paradigms}

\textit{What guiding principles exist for the design of empirical visualization studies, and how might early-career researchers consider them in study design?} Studies are often designed and evaluated from a particular research epistemological stance and based on the type of contribution the authors intend to make \cite{wall_vishikers_2022}. At the highest level, research studies are designed from a particular research paradigm, or beginning from a ``set of interrelated assumptions about the social world'' \cite{ponterotto_qualitative_2005}. The positivism paradigm assumes an objective, knowable reality, and the goal of positivist research is to produce explanations in order to predict or control this reality \cite{park_positivism_2020, ponterotto_qualitative_2005, trochim_2006}. In contrast, constructivist-interpretivist traditions assume that there are multiple realities, each constructed in the mind of the individual. Building on constructivism, the tradition of critical theory believes the constructed lived experience is mediated by power relations, social context, and historical context \cite{ogbonnaya-ogburu_critical_2020, akbaba_shifting_2025, liang_embracing_2021}, and centers the possibility of transformation and the researchers’ values in the research process. Much of theory construction in the social sciences begins from these assumed conceptualizations of reality \cite{jaccard_2020, hawking_2010, trochim_2006, cobley_3_2013}, which researchers then investigate through experimental design. 

To design a visualization experiment and build theory, we explore the idea that for any given study---no matter the epistemological stance or type of contribution---\textbf{researchers construct an experimental space to study people in the real world}. Positivist constructs like external validity lead us to question: \textit{have we constructed a space that allows us to make inferences about the broader population beyond our sample?} \cite{lesko_generalizing_2017, pearl_external_2014} Constructs like ecological validity and mundane realism raise the question: \textit{have we constructed a space that matches the cues people actually experience in a real-world situation?} \cite{kihlstrom_ecological_2021, lu_ecological_2025, orne_social_1962} Ecological validity reminds us to assess how closely constructed environments reflect the everyday contexts in which people typically operate. External validity and ecological validity are subtly different---an experimental space that incorporates all the specificities of real-worldness for a sample can impinge on potential for generalizable inferences to the broader population. 
Constructivist and interpretivist constructs like credibility and plausibility \cite{tracy_qualitative_2010, meyer_criteria_2019} lead us to ask: \textit{why are we constructing a space at all, and removing authentic sources of depth and richness that should be included in our understanding?} Instead, researchers could meet people where they already are and in situations they are already invested in, in order to authentically study varying modes of engagement with visualization. 

We develop this idea of a constructed space further in Sec. \ref{sec:framework}, for our conceptualization of experimental design. We also discuss how interviewees designed studies of the general public from particular epistemological stances in Sec. \ref{sec:exp-vs-real}.

\subsection{Recipes and methods for empirical visualization studies}

\textit{What guiding recipes and methods exist for the design of empirical visualization studies?} Within research paradigms, researchers have articulated multi-stage processes that function as \textbf{recipes or blueprints for visualization studies} of different flavors. Munzner \cite{munzner_nested_2009} puts forth the nested process model to guide visualization system design and evaluation, describing the nested levels of action (domain problem and data characterization, operation and data type abstraction, visual encoding and interaction design, and algorithm design) and threats to validity at each level. Sedlmair et al.’s \cite{sedlmair_design_2012} process model of a design study---in which researchers design and evaluate a visualization system to solve real-world problems of domain experts, and reflect about lessons learned for visualizations broadly---presents a blueprint we thought about adapting for visualization studies that solve problems for broader audiences. McKenna et al.’s \cite{mckenna_design_2014} design activity framework provides guidance on design activities to pursue at different levels of visualization system design, while design decision models like the nested blocks and guidelines model \cite{meyer_nested_2015} specify how to record design outputs and decisions.

Overviews of \textbf{specific methods, or the tools in the researcher toolkit to build out these blueprints}, are also abundant. McKenna et al. \cite{mckenna_design_2014} lays out a smorgasbord of specific design activities to be employed in visualization design to solve real-world problems. Lam et al. \cite{lam_empirical_2012} presents seven evaluation scenarios and evaluation methods applicable to each scenario, synthesized from a literature review of 800 information visualization papers. In an extended synthesis of evaluation methods, Isenberg et al. \cite{isenberg2013systematic} apply Lam et al.'s scenarios to scientific visualization papers.. Carpendale \cite{kerren_evaluating_2008}, Andrews \cite{andrews_evaluation_2008}, and Plaisant \cite{plaisant_challenge_2004} also articulate various evaluation methods in visualization, when to use them, and the purposes and tradeoffs of each. To glean guidance on specific deployment and evaluation methods, researchers can consult write-ups that apply established methods---from user and usability studies to grounded theory \cite{brown_into_2011, kosara_thoughts_2003, zuk_heuristics_2006, faisal_internalization_2008, shneiderman_strategies_2006, saraiya_insight-based_2005, saraiya_insight-based_2006, isenberg_grounded_2008}---to the visualization field.

However, these models presuppose to some extent that a visualization system is the solution, and they cannot help when researchers face the difficulties of finding a common design that works among diverse peoples or attracting engagement from fluid public spaces. Furthermore, though the idea of adapting or proposing a multi-stage process model was enticing, we were wary a process model to guide visualization studies with broader audiences may become too specific (as to only apply to a narrow subset of epistemological stances or research agendas) or too generic (as to apply to everything and not invoke useful, generative brainstorming and reflection).

Instead, we opted to examine a variety of experimental designs of broader, public audiences to learn about similarities and differences in approaches. This led us to develop higher-level conceptualizations and metaphors for empirical visualization study design (Sec. \ref{sec:framework}), over step-by-step recipes, that we additionally pair with interviewees' concrete approaches to experimental design in Sec. \ref{sec:exp-vs-real}. Closer to our paper's contributions is Wall et al.'s \cite{wall_vishikers_2022} actionable breakdown of visualization study components (a research question, study conditions, etc.) and options for each based on the type of contribution the researchers are trying to make (factor, system, or technical).


\section{Investigating recent approaches to visualization for broader audiences}

To understand specific approaches to visualization studies of the public, in addition to our related work, we scoped\footnote{Our process for scoping papers loosely followed the PRISMA process, commonly used in systematic reviews \cite{van_laar_relation_2017, moher_reprintpreferred_2009}.} a set of recent empirical visualization publications for broader audiences and interviewed paper authors to learn about their decisions (Sec. \ref{sec:exp-vs-real}).

\subsection{Paper selection and coding}
\label{sec:methods-paper-selection}

We limited our search to four premier visualization venues: IEEE VIS, IEEE Transactions on Visualization and Computer Graphics (TVCG), Computer Graphics Forum (CGF), and visualization-related papers from the ACM CHI conference on Human Factors in Computing. Any publication that was published or appeared in these venues from 2020 to 2025\footnote{We targeted this time frame to capture studies of COVID-19 visualizations, which became widely familiar to the public, and to improve the likelihood that interviewees could recall their methodological rationales. However, we acknowledge that plans for study deployment were likely impacted by the pandemic. For VIS and TVCG, we only considered publications from 2020 to 2024 because our source \cite{Isenberg:2017:VMC} did not have 2025 works as of download on November 18, 2025.} was considered. 

To filter studies, we compiled a list of keywords starting from topics on the website Our World in Data \cite{data_owid_2024}, which archives visualizations and data related to societal, global issues---a rough proxy for public-facing visualizations. We then added words like ``open data'', ``public'', ``museum'', and ``news media'', as well as words describing general audiences like ``casual user'', ``general public'', or ``non-expert'' (expanding on the search criteria used by Burns et al. \cite{burns_who_2023}). We filtered for the presence of at least one of these keywords in the abstract, title, or paper-related keywords (as well as the word ``visualization'' for papers in CGF and CHI).\footnote{The full list of key words for inclusion and exclusion is included in supplemental materials and on \href{https://github.com/MUCollective/projecting-the-public}{GitHub}.} Publications with the keyword ``virtual reality'', ``augmented reality'', ``vr'', or ``ar'' were excluded, as we were more interested in real-world, public settings.

Querying using just keyword criteria, we initially received \textbf{330 papers}. The first two authors then independently conducted rounds of manual filtering to pick studies that satisfied additional criteria: studies had to concern the impact of a visual or data representation on general audiences and have a means of eliciting responses from participants, inspired by Hullman et al. \cite{hullman_pursuit_2019}. We were interested in visualizations intended to visually communicate meaning versus evaluating skills, and thus excluded studies of visualization literacy testing. Additionally, we excluded studies solely focused on tools or processes for visualization construction, though this should be examined in future research. After resolving disagreements, our final corpus totaled to \textbf{73 studies}.

Inspired by Subramanian et al.’s \cite{subramanian_towards_2023} systematic review of ecological validity guidelines, we coded the 73 papers to understand the landscape of motivations and methods used, and to guide sampling for interviews.\footnote{All 73 papers with corresponding codes are included in supplemental materials and on \href{https://mucollective.github.io/projecting-the-public/}{our study website}.} We coded each study with its methods and measures for the evaluation process and setting for experimental deployment, as well as the visualization's specified audience and role (as a communicative medium, analytic tool, etc.). To ensure we had coverage of categories for interviews, we sampled \textbf{35 studies} through greedy set-cover sampling to maximize category coverage across codes. We looped through all 73 studies and calculated a score for each based on (1) the categories already present in the sample (2) how many new categories across codes the present study would cover if added to the sample. The \texttt{while} loop continues until no study adds new categories, after which additional studies are sampled randomly. The sampling script is included in supplemental materials and on \href{https://github.com/MUCollective/projecting-the-public}{GitHub}.

\begin{figure*}[t]
  \centering
  \includegraphics[width=\textwidth, alt={}]{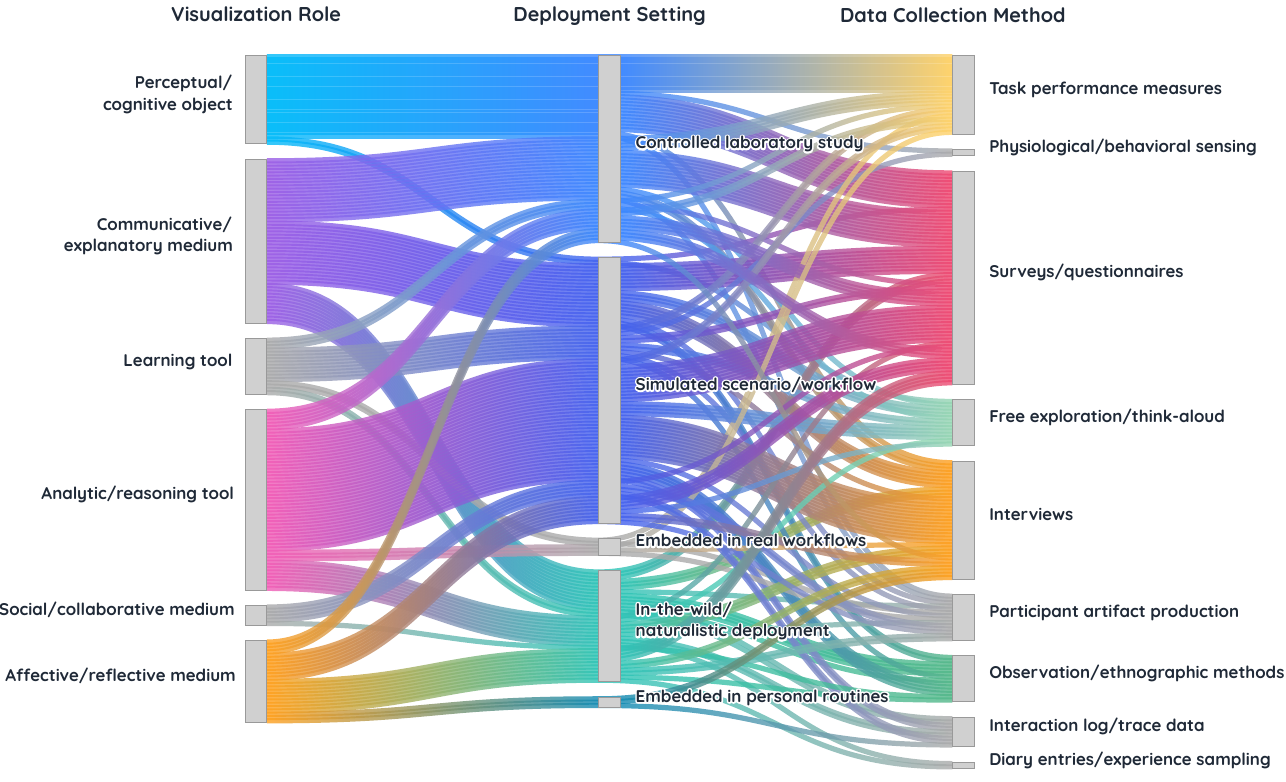}
  \caption{Alluvial diagram showing relationships across the 73 studies among the role of the visualization, the evaluation environment, and the measures used to assess outcomes. Blocks represent categories within each coding dimension, while connecting flows represent combinations of categories observed across studies. Each study is represented by a single strand, and the widths of the blocks and aggregated flows are proportional to the number of studies they contain. Diagram is on our \href{https://mucollective.github.io/projecting-the-public/}{our study website}.}
  \label{fig:alluvial-1}
\end{figure*}

\begin{table}
    \centering
    \tiny
    \begin{tabular}{p{3cm}p{4.625cm}}
        \midrule
        \rowcolor{lightyellow}
        \multicolumn{2}{l}{\textsf{\textbf{Specified audience}}} \\
        \cmidrule(lr){1-2}
        \textsf{General audiences} & \textsf{Specific publics/lived communities} \\
        \midrule
        62 & 11 \\
    \end{tabular}

    \begin{tabular}{p{2.8cm}p{2.2cm}p{2.2cm}}
        \midrule
        \rowcolor{lightyellow}
        \multicolumn{3}{l}{\textsf{\textbf{Specified role of visualization}}} \\
        \cmidrule(lr){1-3}
        \textsf{Communicative/explanatory medium} & 
        \textsf{Analytic/reasoning tool} & 
        \textsf{Affective/reflective medium} \\
        \midrule
        32 & 25 & 15 \\
        \cmidrule(lr){1-3}
        \textsf{Perceptual/cognitive object} & 
        \textsf{Learning tool} & 
        \textsf{Social/collaborative medium} \\
        \midrule
        15 & 5 & 3 \\
    \end{tabular}

    \begin{tabular}{p{2.5cm}p{2.4cm}p{2.4cm}}
        \midrule
        \rowcolor{lightyellow}
        \multicolumn{3}{l}{\textsf{\textbf{Evaluation setting}}} \\
        \cmidrule(lr){1-3}
        \textsf{Simulated scenario/workflow} & 
        \textsf{Controlled lab study} & 
        \textsf{In-the-wild public deployment} \\
        \midrule
        34 & 29 & 11 \\
        \cmidrule(lr){1-3}
        \textsf{Embedded in personal routines} & 
        \textsf{Embedded in real workflows} \\
        \midrule
        2 & 2 \\
    \end{tabular}

    \begin{tabular}{p{2.2cm}p{2.5cm}p{2.6cm}}
        \midrule
        \rowcolor{lightyellow}
        \multicolumn{3}{l}{\textsf{\textbf{Evaluation measures}}} \\
        \cmidrule(lr){1-3}
        \textsf{Surveys/questionnaires} & 
        \textsf{Interviews} & 
        \textsf{Task performance measures} \\
        \midrule
        48 & 26 & 18 \\
        \cmidrule(lr){1-3}
        \textsf{Free exploration/think-aloud} & 
        \textsf{Observation/ethnography} &
        \textsf{Participant artifact production} \\
        \midrule
        10 & 9 & 8 \\
        \cmidrule(lr){1-3}
        \textsf{Interaction log/trace data} & 
        \textsf{Physiological/behavioral sensing} &
        \textsf{Diary entries/experience sampling} \\
        \midrule
        7 & 2 & 1 \\
    \end{tabular}
    
    \caption{Total category counts across publication codes. Publications could be labeled with multiple categories for each code. Definitions of codes and categories, along with the full list of coded publications, are provided in supplemental materials.}
    \label{table:codes}
\end{table}

\subsection{Semi-structured interviews}

\subsubsection{Interview recruitment}

We reached out to the leading authors of the sampled 35 studies to interview them about their study's experimental design, anticipating that a quarter to a third would respond. We also reached out to other visualization researchers who have advised multiple visualization studies with general audiences. Participants consented to the interviews with a consent form approved by our university Institutional Review Board. Our semi-structured interviews---in total, \textbf{15 interviews}---lasted from 60 minutes to 90 minutes, and participants were compensated 20 USD. All 15 interviews were conducted and recorded by the first and second authors over Zoom from January to February 2026.

\subsubsection{Interview format}

In the first block of interview questions, we asked interviewees to describe the motivation behind the specific studies they were contacted about. We then probed for how they determined the definition of ``general audience'' in their visualization study. Given our interest in how researchers construct an experimental space to study real-world phenomena, we also asked interviewees how they selected and simulated the real-world context or use situation for their study.\footnote{While initially we focused interviews around \textit{ecological validity}---how interviewees ensured their findings would meaningfully translate back to real-world conditions---we moved away from that construct to consider broader study influences.} Next, we asked them to discuss other goals or criteria that guided their study design, memorable challenges encountered and tradeoffs made, alternative study designs they considered but ultimately did not pursue, and how they navigated prevailing norms in their research community around experimental design. 

While the interview participant answered questions in the first block, the interviewer recorded key influences to study design mentioned by the participant on virtual Post-it notes in a shared Miro board. These notes were presented to the participant in the second block of the interview, to spark additional editing and additions. We asked participants to sort the Post-it notes (or a subset of notes) from more influential to less influential in the design of their study. ``Influential'' was left to participants' interpretations, as long as they could clarify their definition---some participants defined it as largely affecting the study's practical implementation, as describing factors they thought more about, or as describing what they had more control over.

The interviews concluded with forward-looking questions about guidance the participant would offer to others around designing visualization studies for broader audiences.\footnote{The full interview script is included in supplemental materials.}

\subsubsection{Interview analysis}

We transcribed the interviews with Adobe Premier or Zoom transcription and manually corrected the transcripts. The first and second authors analyzed the transcripts using Braun and Clarke's \cite{braun_reflecting_2019, braun_using_2006} method of reflexive thematic analysis, reviewing each other's analysis of interview chunks to ensure salient and resonant themes in each were captured. 

On analyzing interviewees' transcripts and their Miro board rankings, we saw a persistent pattern of interviewees listing their broad mission and motives for the study first (\textbf{aspiration}), separated from the challenges faced and eventual decisions made to realize the study (\textbf{reality})---which aligned with our personal experiences studying public-facing visualizations. Higher-level themes of aspirations and practical realities guided us toward our 3-part conceptualization (Sec. \ref{sec:framework}), as they aligned with its components and were ultimately mapped to them (in Sec. \ref{sec:exp-vs-real}) once the conceptualization solidified.

\section{Conceptualization of experimental design}
\label{sec:framework}

\begin{quote}
    \textit{The obscure moon, lighting an obscure world.}
    
    ---Caden Cotard, Synecdoche, New York \cite{synecdoche_2008}
\end{quote}

We decided to explore the concept that \textbf{researchers construct an experimental space to study people in the real world \cite{jaccard_2020, hawking_2010, trochim_2006, cobley_3_2013}} within the assumptions of a research paradigm, because this concept was an accommodating foundation from which to compare the diversity of study designs in our corpus. The imagery of a limited space that researchers \textbf{bring people into} communicates \textit{scope}---that studies consider a subset of audiences, that our observations have limits, and that theories have boundary conditions (which describe the generalizability of a theory across contexts \cite{busse_boundary_2017}). In addition to communicating scope, we wanted to make explicit the \textit{involved and disruptive} aspect of constructing an experimental space, wherein researchers build a previously nonexistent space and may ask participants to do certain things by their agenda \cite{yarkoni_generalizability_2022, akbaba_entanglements_2025}. Lastly, from examining visualization studies of general audiences, we aimed to capture the \textit{simplification and focusing} of audience perceptions that occurs when researchers operationalize latent variables into administered measures, reflecting participants' engagement, learning, emotions, trust, and more.


Perhaps a more unconventional approach to developing frameworks, we put pen to paper to brainstorm expressive visual metaphors, shapes \cite{varona_theory_2025}, and philosophical thought experiments \cite{allegory_2026} that could bring these ideas together. We sketched various expressions of a constructed space with these chosen characteristics of experimental design, tweaking them as we conducted our interviews by reflecting on whether interviewees' situations were captured by our metaphors.

Our early drawings represented the experiment as a sandbox or theater set (in which participants see impermanent structures or scenarios for the purpose of the study), a realistic but fictional apartment building (that mimics reality but is ultimately simulated, and that people opt into entering), or a pool that people swim in already (an existing space that researchers attempt to set up a study in). \textit{Scope} was represented as a net that a researcher dipped into a world of people to recruit participants, or a spotlight with fuzzy boundaries cast onto populated buildings. To emphasize the \textit{involved and disruptive} nature of experiments, we illustrated the researcher as a visitor from another world (though researchers can be a member of the community they are studying). And when researchers \textit{simplify or focus} some piece of human experience for measurement, we explored depictions of this as different shadows cast when the spotlight is shone on participants.

Mixing components of our sketches, we propose \textbf{three core elements that illustrate a situational metaphor for designing an empirical visualization study for broader audiences}. What initially proved challenging was creating a metaphor to accommodate different research paradigms and their respective constructs for rigor, but by proposing a situational scaffold rather than step-by-step instructions, researchers have room to adapt its elements. In Sec. \ref{sec:exp-vs-real}, we describe interviewees' aspirations and practical challenges positioned within these elements.

\vspace{5pt}
To design an experiment:
\begin{itemize}
    \item Researchers \iconaudience \textbf{cast audience} \textit{(into a study sample)}
    \item Researchers \iconsetting \textbf{construct experimental setting} \textit{(to study participants in)}
    \item Researchers \iconimpression \textbf{project audience impressions and interactions} \textit{(into participant measures)}
\end{itemize}
\vspace{5pt}


Especially in studies of the general public, researchers need to think carefully about who the audience for the visualization is, and how they \iconaudience \textbf{cast audience} into a study sample. The audience informs visualization design and makes clear who the findings eventually apply to. Like Burns et al. \cite{burns_who_2023}, we believe it important to have clear criteria for the visualization audience and to be transparent about who is eventually recruited in the study. We liken researchers casting audiences to casting a light onto a wider pool of people.

Alongside casting, researchers must figure out how to \iconsetting \textbf{construct an experimental setting} to study participants in that aligns with and supports the study sample and participant measures. 
This setting could be a controlled lab room \cite{sterzik_perception_2023}, an online experimental flow administered through Prolific \cite{padilla_multiple_2022}, or an existing public park that potential participants pass through \cite{morais_exploring_2022}. A constructed experimental setting will always have some degree of artificiality, but attempts to mimic the real world or situate it in real settings can be introduced \cite{kerren_evaluating_2008}.\footnote{Situatedness of setting can be thought of as the degree to which researchers meet and embed in the physical environments where participants are \cite{lam_empirical_2012, twidale_situated_1994}.} The setting includes the visualization or interface shone to participants.


We also envision researchers as \iconimpression \textbf{projecting audience impressions and interactions} into a form they will measure to answer a research question
---the shadow that appears when the light is shined on participants in the setting. Projections can be made through a particular epistemological or critical lens, as we described in Sec. \ref{sec:related-work-paradigms}. In larger, controlled studies, audience impressions and interactions might be projected into a more abstract and isolated mechanism, like in studies of visual perception \cite{sterzik_perception_2023, panavas_juvenile_2022} that incorporate quantitative measures and judgment tasks to understand a generalizable mechanism. Alternatively, they could be mapped into a form that is contextual, subjective, and entangled with factors beyond the visualization shown (that would be treated as confounding factors in a controlled study) \cite{akbaba_entanglements_2025, kauer_discursive_2025, akbaba_troubling_2023} and are often measured through qualitative methods \cite{isenberg_grounded_2008, tracy_qualitative_2010} to capture rich portraits of a person's contextualized internal state. 


We see these elements as a constructive and reflective scaffold to help researchers articulate and interrogate assumptions embedded in experimental design. For junior researchers, the framework provides a vocabulary for identifying design decisions that are often implicit: Who is being cast as the audience? What setting has been constructed to study them? How are audience experiences being projected into measures and evidence? For more experienced researchers, the framework serves as a lens for comparing alternative study designs, exposing tensions between ecological validity and experimental control, or identifying the boundary conditions of claims. The elements can also support mentorship and research communication by providing a common language to discuss how findings emerge from particular audiences, settings, and projections of participant experience.

\section{Applying our conceptualization to a previous study of public election visualizations}
\label{sec:case-study}

We now apply the conceptualization from Sec. \ref{sec:framework} to our previous study of U.S. elections visualizations \cite{cai_2026}. In this study, we grappled with how best to design an experiment to investigate the public's experiences watching visualizations of election results during U.S. elections. The study’s design can be understood through the three elements of our conceptualization as follows:


\begin{itemize}
    \item \textbf{Cast audience:} \textit{U.S. residents recruited from Prolific, of different demographics, geographic locales, and political party affiliations}
    \item \textbf{Constructed experimental setting:} \textit{Participants viewed a live elections dashboard on a web browser, in their personal environments, while votes were counted.}
    \item \textbf{Projected audience experience:} \textit{Participants were instructed to view our dashboard whenever they pleased. They were surveyed and interviewed about their belief in evidence of voter fraud, confidence in the vote count, and overall understanding and experience of our dashboard.}
\end{itemize}

We reflect on instances during this project when we faced real challenges on the road to realizing our aspirations. We aspired to \iconsetting \textbf{construct our experimental setting during the vote count in the 2024 U.S. presidential election}---an uncertain, high-stakes experience that seemed hard to simulate---in order to give credibility to our findings. As part of the realism of our setting, we \iconsetting \textbf{designed a realistic live election dashboard} with real, live-updating vote counts that hundreds of participants viewed on election night. Our rationale for designing our own dashboard, rather than asking participants to view existing dashboards from other sources, was to \textbf{carefully and holistically control the design}---participants saw different variants of our dashboard with new visualizations we were interested in testing. To maximize realism in our design, we consulted our previous design space of news organizations' live elections dashboards \cite{cai_watching_2024}, prototyped over several months, and conducted formative design interviews to solicit feedback from viewers.

A sobering reality not mentioned thus far was the degree to which we, as researchers, \textbf{felt responsible for audiences' experiences during a critical news event}. And because of this felt responsibility, we spent a lot of time \textbf{debating which of our designs seemed unconventional enough to confuse or disturb audiences.} (Plus, had we concentrated on avant-garde designs, our findings would perhaps be less relevant to practitioners.) But we wondered whether the goal of research \textit{is} to test the unconventional and shift the Overton window of visualization design for particular contexts. By sticking more to the basics, did we sacrifice possibilities that could have had more of an effect on our dependent variables?

Though the potential to \iconimpression solicit authentic engagement---by way of showing real vote counts during an election---was enticing, the reality of doing so involved planning carefully for election night in order to \textbf{responsibly shepherd participants' election viewing}. On the day of the election, we were prepared with a schedule of when polls in key swing states closed, ready to live monitor the dashboard and our participants' experiences. Once votes began coming in, we had several friends and lab mates graciously helping with a flood of participant messages while we were busy maintaining the dashboard. We were also \textbf{flexible in our survey administration during vote counting}, because unforeseen events might occur that we would want to survey participants about. To accommodate this, we developed a system that allowed us to flexibly adjust survey timing and notify participants when new surveys were posted.

Looking back, what would we have done differently? We wonder whether it would have been sensible to \textbf{pick only one aspiration, and anchor and align the rest of the study design to it}. In our real-world study of the general public, we strived towards unhinged ambitions for \textit{everything}: the realism of the election setting, the aesthetic and polish of the dashboard visualizations, the varied measures (measuring trust, understanding, etc.) that we projected facets of participants' experiences onto. 

We could have pursued a more \textbf{piecemeal approach, resulting in a suite of studies each anchored by a different aspiration.} Honing in only on the aspiration of authentically capturing participants' impressions of live election dashboards during a real election, we could have deployed a more descriptive, exploratory study in which we capture participants' impressions of existing live election dashboards from other news outlets. If our primary aspiration was to manipulate components of dashboards and observe the associated effects in some isolated, precise way, we could have first run a study in a simulated vote counting scenario, to at least validate our visualizations in a cleaner, controlled scenario before diving into the deep end of a messy, real election deployment. Rather than trying to heavily control an in-the-wild study, or expect rich, authentic engagement in a controlled lab setting, we are \textbf{reminded of the importance of choosing an objective and aligning the elements of study design to that objective rather than trying to have it all}.

\begin{guidebox}
    \guideboxtitle{CONSIDERATIONS \& RECOMMENDATIONS:}
    \begin{itemize}[leftmargin=0.5em, 
                    before=\vspace{3pt},
                    after=\vspace{1pt}]
        \item \textbf{Recognize that testing in popular public settings means participants come in with strong prior expectations of the visualization design.} \textit{Researchers may have to find a midpoint between what participants are used to seeing, and newer, unexpected designs. Especially in high stakes communication settings, researchers should avoid confusing or alienating participants.}
        \item \textbf{Consider designing a variety of adaptable channels of data collection that can be stitched together later.} \textit{In public settings, researchers can benefit from offering multiple ways for participants to contribute, including lower- and higher-effort options as well as scheduled and unscheduled opportunities for engagement. Researchers should be able to adjust the timing of data collection and refine question wording as the study unfolds. It is also helpful to establish a monitoring plan for participant responses once the study goes live.}
        \item \textbf{Consider decomposing ambitious research agendas into multiple complementary studies.} \textit{Rather than expecting a single study to satisfy competing goals, conduct a suite of studies that each prioritize different aspirations. For example, controlled studies can isolate causal mechanisms, while in-the-wild deployments can capture authentic audience experiences and generate hypotheses in an unexplored space.}
    \end{itemize}
\end{guidebox}

\section{Aspirations vs reality: Why visualization studies of broader audiences are hard to design}
\label{sec:exp-vs-real}

Now, we turn to the experiences shared by our interviewees. Findings are organized as interviewees' aspirations met with realities of designing and conducting studies for general audiences, that we pinned to an element of the 3-part conceptualization. Actionable guidance from findings are then summarized below each section.



\vspace{1mm}
\subsection{When casting audience}
\label{sec:exp-vs-real-audience}

\paragraph{\textbf{\iconaspiration \iconaudience Researchers cast audience widely to capture a broad and diverse set of perspectives in the study sample.}}

When studying broader audiences, interviewees (P2, 6, 8, 10, 14) discussed the hope of \textbf{recruiting a wide diversity of perspectives through different mediums}---of zooming out and shining a light on a broad swath of people, or by shining several lights on multiple populations. There was a sense that surveying a large, diverse population was an opportunity (P1) because of the possibility for a visualization to have far-reaching impacts (P9, 10). Interviewees endeavored to consider and design for everyone who \textit{did not} have a high level of visualization or data expertise (P3), or anyone who might be interested in a publicly-accessible tool (P4).

\paragraph{\textbf{\iconreality \iconaudience Casting audience widely makes it difficult to pick a visualization design and hypothesize about its effects.}}

However, casting widely for a diversity of perspectives \textbf{creates headaches for deciding what the visualization stimuli or intervention should even be} (falls under \iconsetting construct experimental setting)---how do we make hypotheses about the effect of our visualization, or what the ``right'' visualization is, in a sea of diverse public opinions? As P4 noted, \textit{``We had actually quite some problems to define the target audience, because constantly people kept telling us, or also my supervisors kept telling me, this is way too broad.''} 

One approach is \textbf{falling back to accepted heuristics} (P1) in visualization design that have been shown to achieve a certain effect, like adding personalization to data to engage viewers. Interviewees \textbf{prioritized simplicity and aesthetics} (P2, 3, 5, 6, 9, 10, 13) in their visualizations by distilling complex data into a narrative with a few clear messages, or focusing on visual clarity. Imagining some simple baseline of common knowledge across all participants, P2 and 13 targeted visualization designs for an elementary or high school reading level and assumed that audiences would know enough with that lower threshold of schooling to engage.

But is falling back to general heuristics or simplified language always the right choice? Trying to consider every perspective can mean relying on a composited silhouette of a ``general audience'' that means different things to different people---P14 discussed how targeting general audiences can involve \textit{``just a lot of projecting on the part of the researcher''} and \textit{``there's this risk that we're just, you know, designing for ourselves, and thus not generating something that's representative.''} Trying to consider every perspective can, regrettably, result in considering no one in particular. Furthermore, there was recognition (P1, 7, 9, 12, 15) that not all individuals will encounter the visualization given that studies usually evaluate a biased slice of the population through convenience sampling.

As an alternative approach, interviewees (P11, 14) \textbf{considered smaller pockets of people that identify by shared, lived experiences}, but who still varied greatly in their knowledge of data and visualization. Informed by a more interpretivist perspective, P14 focused on smaller groups and discussed how diversity of perspective can become apparent the closer one examines a seemingly niche group of people. Rather than shining the audience light broadly, P14 stated, \textit{``... and so you have to just kind of keep drawing smaller and smaller boxes} [or in our metaphor, focusing the light on smaller areas with better illumination] \textit{before I start to feel comfortable actually making any sort of meaningful claim about the set of abilities or experiences or needs of a set of people.''} This suggests an orientation to visualization design and evaluation centered more on understanding niche, lived communities.
 
Yet another approach is to think about diversity as broadening the primary audiences represented by visualization research---P1, 11, 12, 14, and 15 \textbf{sought to reach populations that researchers normally do not capture}. For example, P15 set out to study bilingual populations who are underrepresented in visualization research, and P11 surveyed individuals in rural U.S. communities excluded from the countless visualization studies of university populations. To do this, a considerable amount of thought and coordination was put into recruitment strategies. P15 stated that she \textit{``recruited from a set of, like, four universities, so 2 were in India, and two were in, like, Dubai''} and P11 had to \textit{``completely flip the model''} of recruitment and \textit{``went specifically to places where we felt like was a better representation of the population in the area, and where they were already there, and already engaged.''}

Still, interviewees (P1, 12, 14) talked about how some visualization effects do generalize over a large enough number of people at some abstract level, that allows you to \textit{``kind of at some level, like, smooth over a lot of those differences, and hope that you can do some sort of design work that will hit at least a large or interesting set of them''} (P14). For studies examining how a visual variable performs, P1 described the hypothetical scenario of recruiting 1,000 people over Amazon Mechanical Turk \textit{``that don't have anything else that is specific about them''} and who are \textit{``just human beings that look at the charts, and I think that's totally fine.''} From this perspective of prioritizing larger, diverse sample sizes and generalizability, P12 \textbf{conducted multiple visualization studies across several hundreds of participants recruited online}, testing different data sets, colors, annotations, topics, and chart types in the stimuli. P12 discussed their approach as attempting to get closer to the experimental ideal of generalizability, while acknowledging the bias of online populations.

No matter the chosen approach, P8, 9, and 11 emphasized being clear about chosen definitions of ``general audience'' by \textbf{including the criteria for inclusion in the study and where researchers are recruiting from} (lest we use a term that means something different to everyone).

\begin{guidebox}
    \guideboxtitle{CONSIDERATIONS \& RECOMMENDATIONS:}
    \begin{itemize}[leftmargin=0.25em,
                    before=\vspace{0.5pt}]
        \item \textbf{Choose a recruitment strategy that aligns with the intended research agenda.} \textit{Researchers should prioritize large sample sizes if they intend to identify robust aggregate effects, but focus on a smaller subset of a lived community if they want to paint a saturated picture of visualization engagement. Researchers should
        narrow their target audience and contexts when making stronger claims about the needs, abilities, or experiences of people.}
        \item \textbf{Specify how the ``general audience'' is being constructed.} \textit{Terms such as ``general public'' can give the impression of a homogenous, all-encompassing participant pool despite substantial variation and limitations in recruitment, so being explicit is paramount.}
        \item \textbf{Invest in reaching populations who are routinely excluded from visualization research.} \textit{Allocate time, funding, and effort toward community partnerships, multilingual materials, and recruitment practices tailored to the contexts in which excluded populations already gather, to make participation easier.}
        \item \textbf{Recognize that designing for a ``general audience'' often involves assumptions and projections.} \textit{Critically examine whether visualization design decisions are grounded in evidence about intended audiences, or are instead based on accepted heuristics and intuitions. Challenging existing defaults can be a research project in it of itself.}
    \end{itemize}
\end{guidebox}

\paragraph{\textbf{\iconaspiration \iconaudience Researchers cast audiences in public spaces in the moment to document organic interactions.}}

When interviewees wanted to study authentic engagement with visualizations (falls under \iconimpression project audience impressions), they aspired to cast audiences in public spaces in real time, after individuals had decided to \textbf{engage with deployed visualizations of their own volition and interest}. This aspiration for recruiting in public spaces is tightly coupled with interviewees' objectives to research \iconsetting particular settings, like museum settings (P2, 13) that offer an informal, multi-user learning environment with visualization exhibits.

\paragraph{\textbf{\iconreality \iconaudience Casting audiences in public spaces introduces fluidity and uncertainty in the makeup of the study sample.}}

In publicly accessible spaces, like museums or on social media, an endless variety of people \textit{could} feasibly encounter the visualization under study---there is a high level of uncertainty as to who comes through and engages with a visualization in these public spaces. Unlike in studies with more control over the selected participant pool, it was \textbf{difficult to guarantee any kind of balanced sample} using fluid, in-the-moment recruitment strategies. P2 conducted \textit{``observation sessions at different times of day and different days of week''} but stated that, \textit{``You just, you know, there were some days I show up, sit there for two hours and no one, like, almost no one would come, right? And how was I supposed to know?''} However, P2 opted to prioritize casting audiences through the organic, naturalistic ways they engaged with the visualizations, even though a perfect study sample was not guaranteed. 

Interviewees (P1, 9) discussed the reality that the most interested and opinionated would ultimately participate in a study, analogous to situations in which the most displeased customers leave reviews for a restaurant. Describing respondents who engaged with visualizations distributed through a newsroom, P1 said, \textit{``So they, you know, if you take time to fill out a survey, it means that you care. And I think that already kind of skews it partially right, because you are already giving up your time for that.''} P1 mentioned people who only participated because they saw the visualization and \textit{``want to tell them how much I hate them, so I will take the survey and do the survey and be very opinionated about it.''} They received a lot of survey replies that \textit{``either were, oh, this is cool, love it, or was it like, why did you do it, this is horrible, it's not working at all''} and calling their materials \textit{``leftist propaganda.''}

In reality, \textbf{changes to the recruitment might be enacted while it is ongoing in the public space.} For example, P6 initially planned to \textit{``wait for people to just stop by and get interested in the visualization, and then maybe asking questions, or trying to figure the visualization out by themselves.''} But after a couple days, they realized they would not achieve an adequate sample size, and P6 discussed how they began to initiate interactions: \textit{``People were, like, jogging or walking their dogs, and we would ... we were, like, crazy. Hey, hello, how are you? Well, would you like to, to look at this thing? What do you think?''} By making this change, they traded off the ability to claim that passersby's engagement was purely organic. To avoid this, P13 talked at length about \textbf{designing audience hooks} for their visualization museum exhibit, using dynamic animation or tactile objects to draw in museum visitors, because the \textit{``public has to have a certain level of hook and narrative in order for them to actually buy into it.''}

Even with the uncertainty and real-time changes in participation, it is still possible to roughly predict audience characteristics---P13 was \textbf{aware of the overall audience demographic that their museum partner served} and thus who might see their exhibit. And because there will still be aspects of recruitment that elude control and certainty, interviewees (P1, 2, 9, 12) simply \textbf{acknowledged the self-selection bias in their participant pools} as a tradeoff of exploring organic recruitment methods. Deploying widely in public spaces also opens studies to responses that vary in sentiment and relevance to the research project (falls under \iconimpression project audience impressions), as experienced by P1. To help with this, P10 recommended that researchers \textbf{take quality checks for responses seriously} when conducting studies with the broader public.

\begin{guidebox}
    \guideboxtitle{CONSIDERATIONS \& RECOMMENDATIONS}
    \begin{itemize}[leftmargin=0.25em,
                    before=\vspace{0.5pt}]
        \item \textbf{Choose a recruitment strategy that aligns with the intended research agenda (again).} \textit{If the goal is to understand how people naturally encounter and engage with visualizations, researchers should recruit in spaces where those interactions already occur. However, lack of participation and self-selection bias are tradeoffs of organic recruitment that can compromise planned sample sizes.}
        \item \textbf{Design audience hooks intentionally to encourage voluntary engagement.} \textit{When relying on voluntary participation, visualizations may need compelling entry points, like animation or physical artifacts, that encourage engagement.}
        \item \textbf{Document and justify any in-the-moment changes to recruitment procedures during deployment.} \textit{Public studies may require adjustments after deployment, such as moving from passive observation to active recruitment. Designing studies to be adaptable also means designing studies to be documentable and traceable.}
        \item \textbf{Prepare for a wide range of responses and implement response quality checks for public-facing studies.} \textit{Open public deployments can attract highly enthusiastic participants, highly critical participants, or off-topic responses, so researchers need to think through which responses to report based on research goals.}
    \end{itemize}
\end{guidebox}

\subsection{When constructing experimental setting}
\label{sec:exp-vs-real-setting}

\begin{figure}[t]
    \centering

    \includegraphics[width=\columnwidth]{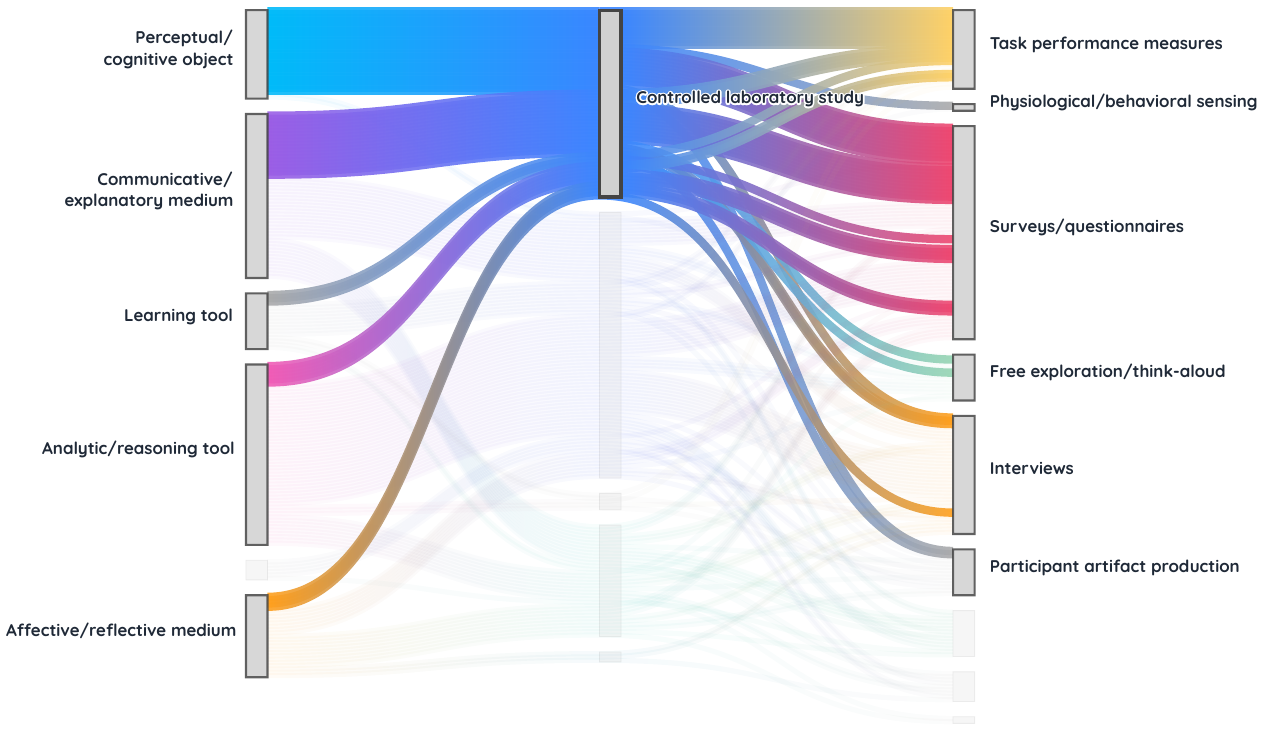}
    \vspace{4pt}

    \includegraphics[width=\columnwidth]{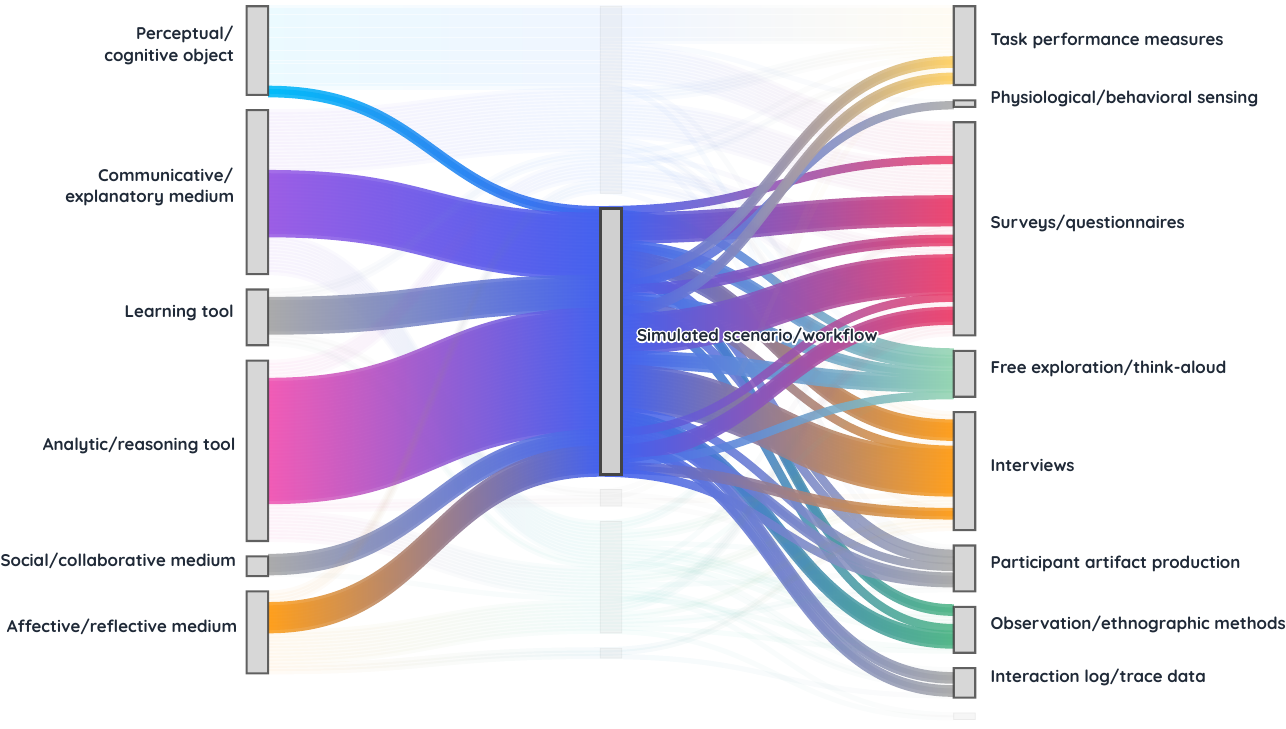}
    \vspace{4pt}

    \includegraphics[width=\columnwidth]{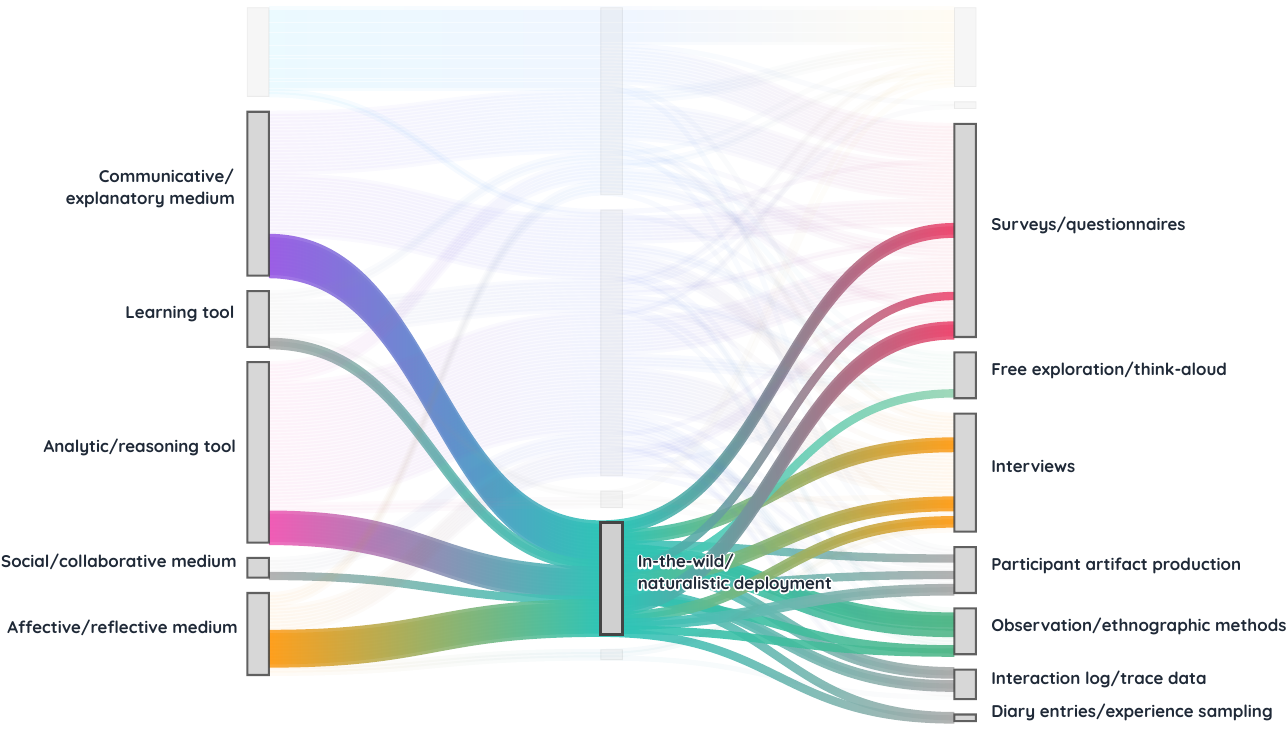}

    \caption{Studies filtered by deployment setting.}
    \label{fig:sub-alluvial-1}
\end{figure}





\paragraph{\textbf{\iconaspiration \iconsetting Researchers construct realistic settings to lend credibility to findings and ensure relevance among participants}}

Using real-world stimuli, scenarios, and surroundings was important to some interviewees (P3, 5, 7) because it \textbf{made findings more convincing}. There was a concern among interviewees of researchers forcing an effect in audience responses (falls under \iconimpression project audience impressions) with synthetic study settings and stimuli that otherwise would not occur in the real world. Regarding studies of visualization deception, P3 discussed how previous experiments used artificial charts with systematically applied distortions, whereas \textit{``real deceptive charts on the web, don't follow this, you know, neat, you know, experimental, like paradigm, where all of the y-axes are inverted a certain way, or all of the, you know, slopes are compressed a certain way.''} In the same vein, P5 talked about finding real-world datasets to demonstrate their visualization tool rather than designing a dataset to be \textit{``the exact scenario or exact dataset that you're envisioning''} (like fitting a model to its training data to demonstrate its success).

Interviewees (P5, 10) also spoke about real-world settings as \textbf{affording an immediate relevance to participants}. When determining the experimental context or scenario, P5 discussed this perceived relevance as necessary in order make a strong case about the usability of the tool, considering, \textit{``What is the context, that even if the context [...] is not present in their life, can they connect to it? Can they think that this might be a problem in the future, or this might be relevant for us?''} And to ensure relevance and interest to participants, P10 mentioned how they run \textit{``a lot of these pre-studies and figure out what is the data that is most interesting.''}

\paragraph{\textbf{\iconaspiration \iconsetting Researchers construct realistic settings to derive new insights into visualization engagement grounded in participants’ existing behaviors.}}

By starting from in-the-wild visualizations with existing relevance to the public (P1, 8, 9, 11)---for example, rainbow maps (P1) or weather visualizations from weatherfluencers (P11)---interviewees talked about the possibilities for \textbf{discovering alternative models of visualization design and engagement} that \textit{``breaks a lot of, like, our computer science values''} (P11). As an example, the inspiration for P8's interview study stemmed from a real-world observation of designers employing certain visualization styles (despite accepted conventions that deem these styles wrong or misleading). During interviews, they showed real-world examples of this style to understand when and why participants preferred them.

\paragraph{\textbf{\iconreality \iconsetting Constructing complex, realistic settings makes it difficult to determine experimental conditions and isolate an effect.}}

Because real-world visualizations can be complex (P3, 8, 9), it was \textbf{difficult to consistently apply an experimental manipulation} across real-world examples. For example, P3 had to \textit{``throw out some pretty, like, technical and complicated charts''} as stimuli because the planned experimental manipulation would have confused the meaning of the real-world charts for participants. In some cases, it was also difficult to study any one characteristic of real-world visualization designs (that would have an effect in isolation) because of how much the examples varied from one another---during a study of engagement with real-world data physicalizations, P6 highlighted the reality that \textit{``the design aspects, the physical encodings, or visual encodings of the visualizations, or how we framed our discourse, they're completely different.''}

Using more complex, niche, real-world visualizations and scenarios also \textbf{complicates finding a baseline condition or benchmark} (P1, 4) ---what is a reasonable, ``neutral'' benchmark that is not so different that it represents a completely different context? Rather than investigating whether their deployed visualization was better than some benchmark, some interviewees (P1, 4) specified their study intent as \textbf{identifying what design approaches could} \textit{``cause a change in how people perceive the data and how much people spend time with the data''} (P1) and to first \textit{``establish this} [more niche] \textit{medium''} (P4).

An approach that combines realism with the ability to isolate effects from aspects of visualization, researchers might begin with a study \textit{``as real as we can do without, being, having, wickedly large number of confounding variables''} (P12), but \textbf{gradually move to more rigorously controlled studies that titrate out other factors} to understand the effect of a single characteristic. They could also, as P15 and Stokes et al. \cite{stokes_striking_2022} did, \textbf{strip visualizations down to their base components}---what is minimally needed for a line chart, for example---and then \textbf{systematically layer other components}, like annotations and labels, to investigate their effects in isolation and in combination with each other.

\begin{guidebox}
    \guideboxtitle{CONSIDERATIONS \& RECOMMENDATIONS}
    \begin{itemize}[leftmargin=0.25em,
                    before=\vspace{0.5pt}]
        \item \textbf{Anticipate that starting from real-world stimuli and scenarios may make clean experimental manipulations and baseline conditions harder to define.} \textit{Complex real-world examples often vary in ways that make it difficult to isolate a single design factor. Researchers can frame their work as more exploratory, and emphasize investigating the effectiveness of several existing real-world methods that vary across dimensions.}
        \item \textbf{Progress from realistic studies to more controlled studies when seeking to isolate causal mechanisms.} \textit{Realistic deployments can clarify which visualization factors are worth pursuing. Researchers should then follow up with more controlled experiments that decompose visualizations into core components, with features  systematically added, to study and isolate effects.} 
    \end{itemize}
\end{guidebox}

\paragraph{\textbf{\iconaspiration \iconsetting Researchers construct situated settings embedded in participants' physical environments to observe engagement in context.}}

Similar to the reasoning for \iconaudience casting audience live in public spaces, P2, 6, and 13 performed in-situ studies to \textbf{study engagement with visualizations in context}, as they relate to the immediate social and physical environment. P2, 4, and 13 emphasized the impact of environment on participant mindset and how they engaged with a visualization, arguing that even subtle attributes of an environment can impact interactions with visualization (P2).


\paragraph{\textbf{\iconreality \iconsetting Constructing a situated setting requires compromise in study deployment and collaboration with institutional partners.}}

One consequence, however, of conducting research in natural settings is having to settle with a less than ideal deployment setting, if the space hosted by the collaborator has its own goals and constraints that prevent researchers from displaying the visualization as intended. P2 described an existing museum room that would have been a good fit for their visualization exhibit, but stated that \textit{``... we couldn't actually put our thing in there. And so instead, we put it kind of just outside. But the thing about it just outside is that it was kind of this little hub in between all the rooms with exhibits, and it was kind of like beside bathrooms and stuff. Like, it was kind of an odd space.''} P2 navigated this by \textbf{treating both ideal and less ideal deployment settings as distinct study conditions}, creating an opportunity to examine how different contexts shaped visitor engagement.

\textbf{If there is no room for compromise, then it may be worth running the study in a controlled environment}. To avoid added uncertainty and constraints from situated settings, P7 conducted their study in an indoor lab setting with optimal lighting conditions, specifying that their primary goal was finding the optimal parameters and that conditions in the wild might be \textit{``too uncertain''}. Simulated conditions may also be required when there are safety risks involved in a situated field deployment. P10 described a study testing wearable visualizations for use while driving as, \textit{``a very particular situation where I'm not even sure I would want to do a field deployment, because imagine I'm building a [...], and I'm sending people off with their car to drive around, and they look at their thing, and then drive into a ditch, because, you know, this, I think this would ethically, would be very, very difficult.''}

To set up in existing settings and conduct more situated research, interviewees (P1, 2, 13, 14) \textbf{fostered relationships and trust with institutional collaborators} to gain access to these public spaces. Researchers may develop long-term relationships with collaborators prior to conducting a study, as P1 did.

\begin{guidebox}
    \guideboxtitle{CONSIDERATIONS \& RECOMMENDATIONS}
    \begin{itemize}[leftmargin=0.25em,
                    before=\vspace{0.5pt}]
        \item \textbf{If deploying in a public space, take advantage of unexpected deployment environments as potential study conditions.} \textit{When partnering with a public institution, rather than viewing unplanned deployment locations as methodological failures, researchers should leverage this to understand how variation in physical contexts influences engagement.}
        \item \textbf{Use controlled settings when environmental uncertainty would compromise study objectives, or when there are ethical concerns.} \textit{However, controlled environments are preferable when researchers need to minimize confounding factors, or when field deployments would create safety hazards.}
    \end{itemize}
\end{guidebox}


\subsection{When projecting audience impressions and interactions}
\label{sec:exp-vs-real-experience}

\begin{figure}[t]
    \centering
    
    \includegraphics[width=\columnwidth]{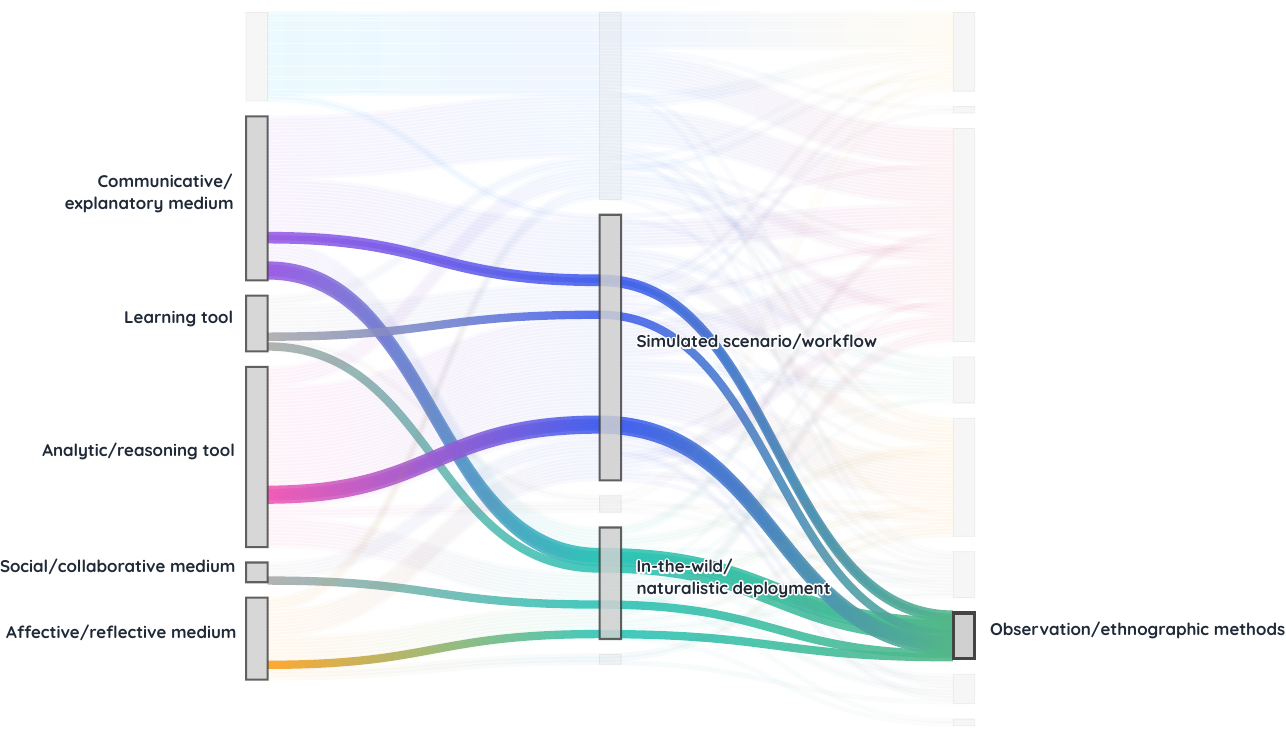}
    \vspace{4pt}

    \includegraphics[width=\columnwidth]{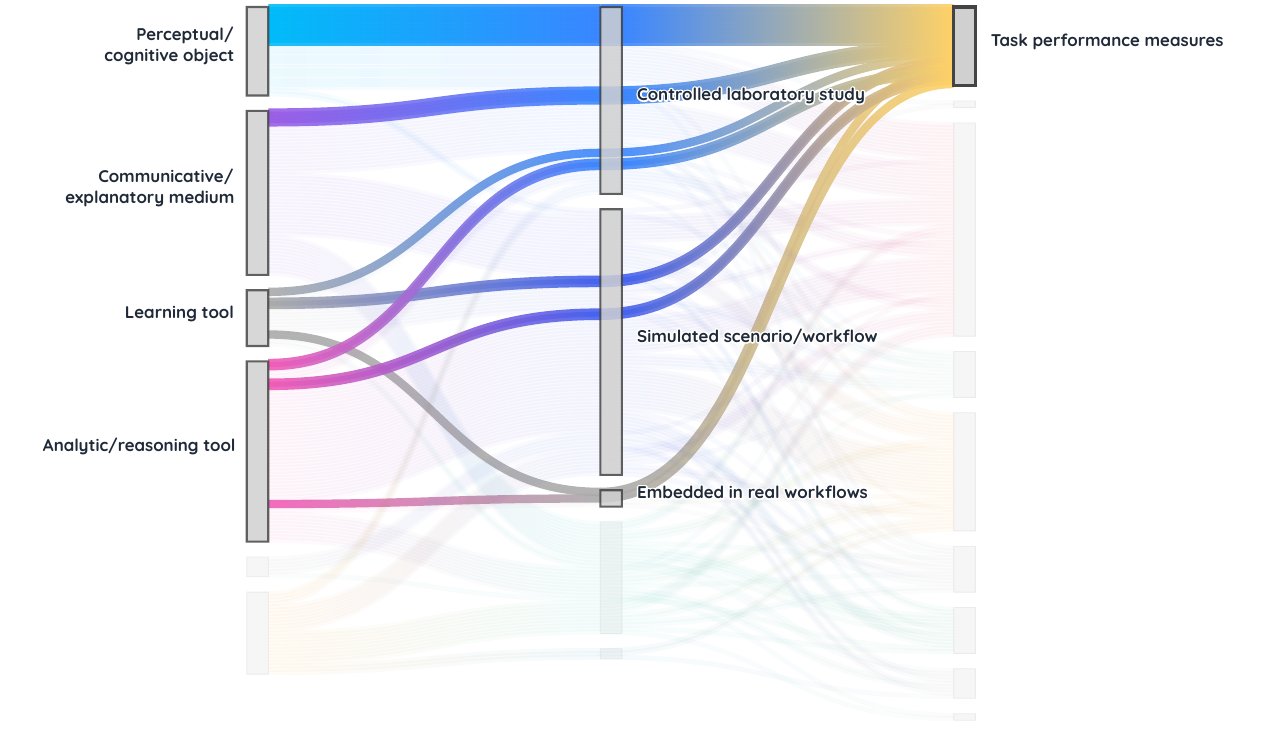}
    \vspace{4pt}

    \includegraphics[width=\columnwidth]{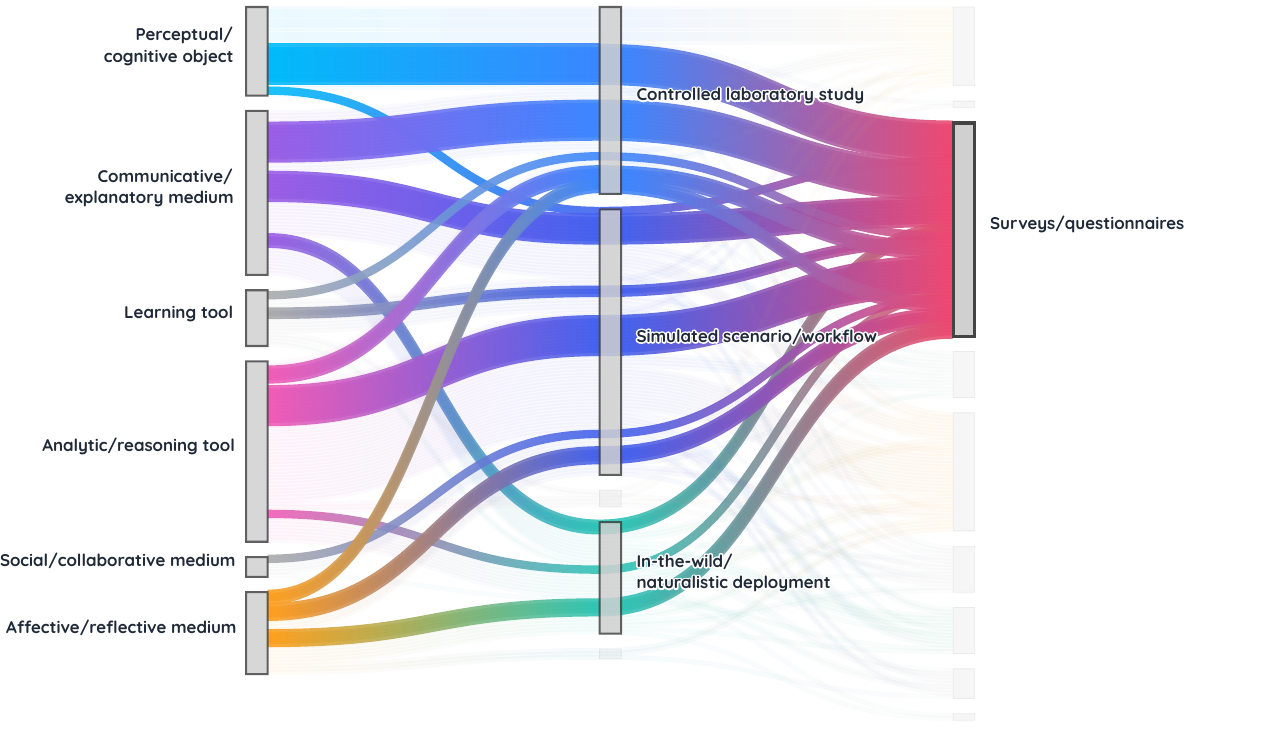}
    \vspace{4pt}

    \caption{Studies by method for collecting participant data.}
    \label{fig:sub-alluvial-2}
\end{figure}


\paragraph{\textbf{\iconaspiration \iconimpression Researchers assess the effects of visualization design (and the underlying data) on participants to produce research of interest to the visualization community.}}

Interviewees discussed the visualization field as \textbf{caring about questions and implications for visualization design and performance} (P1, 9), and \textbf{professional, domain-specific contexts over personal contexts} (P15). P1 commented that, \textit{``at least from my perspective, this, 99\% of the conference is computer scientists that come from very scientific domains, that bring forward papers that are, I improved by 0.01\% how legible this bar chart is in showcasing, like, how this protein binds to a certain receptor in the body for a biology ...''}  P9, reflecting on a study around political visualizations, said, \textit{``I didn't expect the vis research community to really care about that one way or the other [...] Because [...] that's political science. It's not, it's not really, like, a, like, a design question, I think.''} This translated into a desire among interviewees to solicit participants' responses (and project a shadow)---however minimal---regarding how they were affected by the dataset and visual encodings (and the information they derived from the visualization) in order to \textbf{produce research relevant to the visualization community}.

\paragraph{\textbf{\iconreality \iconimpression External factors, like personal experiences and domain knowledge, muddy (and enrich) participant responses.}}

Because broader audiences have a wide variety of backgrounds, motivations, and interests entering a visualization study (Sec. \ref{sec:related-work-broader}), their reactions may \textbf{extend beyond the immediate visual encodings and data (and what researchers have projected for measurement)}. Especially when \iconaudience casting widely in public spaces and \iconsetting constructing a setting with real-world examples, engagement with the visualization can go beyond thoughts about the positioning of tick marks or opacity of bars.

The visualization design or topic may trigger emotions or personal experiences---P8 detailed how their participants introduced personal experiences: \textit{``I had, like, one participant [...]. And he had a lot to say about this, and he was linking it to himself and his family. That was helpful, but also that was, like, for at least 3 minutes, talking about [...] how this is related to his personal life. [...] And they get attracted to those that speak about them, speak about their cities, speak about their financial situation, so on.''} To get the best of both worlds, P14 utilized visualization cards during interviews to \textbf{ground the conversation in data and visualization, but still allow people freedom to discuss their own experiences} related to them.

Interviewees (P3, 9) were aware that real-world charts could trigger strong reactions before the participant really engaged with the new design or tool. In using more political charts, P9 \textbf{considered the political priors that participants arrived with}: \textit{``So, like, with this study specifically, like, it [...] was a huge challenge, because, it's all, like, a lot of this stuff is based on, is so contingent on your priors, your political priors [...] And it's, it's very hard to move the needle on, like, already entrenched ideas, where people kind of know the party script.''} To deal with this, P3, 9, and 15 suggested \textbf{realistic visualizations on ostensibly ``neutral'' topics}---\textit{``the price of milk or the price of oil, as I said, over time''} (P3), \textit{``kind of more toy topics''} (P9), \textit{``nothing climate-wise, nothing political, nothing, like, you know, abortion rights, like, nothing that can remotely be construed as, like, a value morality kind of thing''} (P15). And perhaps this is appropriate when the researcher's goal \textit{``is to do something else, your goal is to understand how a certain visual variable performs in an A/B test''} (P1) and they need datasets that \textit{``are factual and neutral and you can just showcase the data''} (P1).

To focus responses into the desired projection, researchers could \textbf{administer some amount of priming, scenario-setting, and training} before collecting data from participants, to ensure participants understand the goal of the study. With training, P13 noted not wanting to bias participants too much---though, as P9 stated, \textit{``it's not, like, intrinsically risky, as long as, like, there's kind of a recognition of you are [...] moving your, like, magnifying glass to different parts of the problem as you manipulate different, as you, like, force different scenarios.''} Related to concerns for \iconaudience casting audience, P2, 10, and 13 questioned how to design training that accommodates the varying levels of knowledge and areas of expertise among diverse participants. It may also be impractical to train every passerby for visualizations deployed in public settings (P13). 

Several interviewees (P3, 4, 7, 9, 10, 12, 13, 14, 15) advised researchers to take the time to \textbf{closely understand their audiences and their context} (which could constitute an entire study) and to \textbf{conduct several pilot studies}. As a starting point for study planning, P6, 8, and 11 talked about how \textbf{projecting and measuring audience experiences without clear bounds} can be better for \textbf{when researchers are uncertain about what audiences care about or what the important research question is}. These unbounded, exploratory studies can produce valuable insights in their own right, and serve as a first step to inform future controlled studies of particular and precise mechanisms.

\begin{guidebox}
    \guideboxtitle{CONSIDERATIONS \& RECOMMENDATIONS}
    \begin{itemize}[leftmargin=0.5em,
                    before=\vspace{1pt},
                    after=\vspace{1pt}]
        \item \textbf{Anticipate that participant responses may extend beyond the visualization itself.} \textit{Participants often draw on personal experiences, emotions, domain knowledge, and prior beliefs when interpreting visualizations. It is tempting to ignore them because they complicate causal claims about design, but they provide valuable insight into how visualizations are experienced in practice. Plan ahead as to how out-of-scope responses will be handled.}
        \item \textbf{Consider neutral topics when seeking to isolate the effects of visualization design.} \textit{Less contentious topics can reduce the influence of strong prior beliefs, making effects from visual manipulations easier to claim. Adopt priming, scenario-setting, and training to focus participant attention, but acknowledge this also decreases ecological validity.}
        \item \textbf{When in doubt, conduct pilot studies to refine study materials and measures.} \textit{Pilot studies can reveal unexpected interpretations, misunderstandings, or sources of variation in participant responses before a larger, more costly deployment.}
    \end{itemize}
\end{guidebox}





\section{Discussion}

With our conceptualization in Sec. \ref{sec:framework}, we ultimately wanted to drive home the idea that researchers construct a space when they design a study. Individuals in the general public swim along the mortal coil and go about their day-to-day routines, and when they consent to participating in a study, they enter a different space that researchers have constructed that has varying degrees of realism, situatedness, and researcher control. As a reminder, P2 and 9 advised researchers to look at the bigger picture beyond the immediate visualization interface and to consider how visualizations are seen in the real world outside of the study.

One of our main takeaways from interviews is the reality of working within constraints in visualization studies with broader audiences, to measure some piece of audience experience across what feels like a vast, unending public. Beyond the realities already discussed, there are always considerations of funding and resources (P4, 13, 14) and reviews from university research ethics boards (P4, 10). There are deadlines researchers need to meet (P5). Collaborators might need to leave midway through the study (P13). These matters are familiar to any researcher, not just those who conduct studies aimed at the broader public. The truth may be that, \textit{``there's no answer to this, and we have our ideals, and we have what we want and what we wish could happen, but I think the reality is that the research infrastructure is just not there for us to [...] to match our ideals''} (P12).


\subsection{How can visualization research accommodate broader audiences?}

In scoping papers for our corpus from the chosen visualization venues (Sec. \ref{sec:methods-paper-selection})---albeit through one imperfect approach to scoping---only a small slice (2.1\%) of the queried visualization papers published in the last five years were included. P1, 11, and 15 discussed visualization research's proclivity for professional, scientific, and lab domains, with methods influenced by norms in psychological research (P11). If researchers are to better align research goals with the general public's everyday life and communicative contexts, P1 and P11 discussed a rethinking of the scope of the visualization field---to perhaps consider visualization as cultural artifacts tied into cultural and social issues, or as a form of communication that can be weaponized (P1). Interviewees (P6, 11, 13) stated that this may require a broader acceptance of messier, contextual research with qualitative methods that require researchers to cede the control they have in controlled lab studies. 

Consider the following scenario from P11 and 14: if a study finds that external factors---outside of the visualization design---are what participants care about more, do we abandon the study, or do we report back to the visualization community that the visualization was not the priority?  P11 raised the possibility of expanding the scope of the visualization field from visualization to the larger theme of human data communication, describing the disagreements in the visualization community about research scope: \textit{``There's some people who think it should be just, like, human data interaction. There's, like, this kind of, like, sub-movement in the field that think it should be larger. And there are other people who say that that's not appropriate, we gotta focus in on vis.''}

Moving forward, P11 also advised visualizations researchers to think about the benefit of research to the population(s) studied---what are the short-term, mid-term, and long-term benefits? In an ideal world, visualization tools should be useful for purposes outside of academic publication (P11), potentially embodying the ideals of action research (P1) or reciprocal research.




\section{Limitations}

The studies in our review are limited to the 2020 to 2025 time frame (in which study deployments were likely impacted by pandemic restrictions), and we hope to continue expanding our corpus of visualization studies for general audiences to examine how visualization roles, deployment settings, and data collection methods shift over time. We also acknowledge that our interviews provide a non-exhaustive account that skews toward Western perspectives. Incorporating more voices from the Global South, including researchers and practitioners who have studied visualizations in both academic and non-academic contexts, would be especially valuable. Future work could also examine each assumed role of visualization in greater depth to understand why visualizations designed for different purposes may require distinct methodological approaches to study. 

\section{Conclusion}

In this paper, we reviewed 73 visualization studies that did not specify a focus on domain or data experts, interviewed 15 study authors, and reflected on our own experiences constructing visualization experiments for the general public. These studies expand outside of academic and professional contexts and data analysis tools for domain experts by looking at a range of roles for visualizations, settings for study deployment and evaluation, and participant measures. Our unabashed hope for future visualization research is that it embodies the spirit of designing and creating for broader audiences---designing for a diversity of experiences and expertise, understanding a variety of real-world contexts for visualization in which audiences are already passing through and engaging, and considering underrepresented audiences---while remaining attentive to and transparent about the practical limits of any individual study.



\subsection{Acknowledgments}
\noindent The authors wish to thank the MU Collective lab for their gracious feedback and support on this project, and Benjamin Felix for his beautiful illustrations and willingness to iterate on these ideas.

\bibliographystyle{IEEEtran}
\bibliography{main}

@misc{viscomm_2025,
	title = {VisComm 2025},
    author = {Ottley, Alvitta and Schwabish, Jonathan and Parsons, Paul and Baigelenov, Ali and Rahman, Md Dilshadur and Shukla, Prakash and Pandey, Saugat},
	url = {https://www.viscomm.io/},
	language = {en},
	year = {2025},
    lastaccessed ={March 16, 2026},
}

@article{wang_emotional_2019,
	title = {An {Emotional} {Response} to the {Value} of {Visualization}},
	volume = {39},
	copyright = {https://ieeexplore.ieee.org/Xplorehelp/downloads/license-information/IEEE.html},
	issn = {0272-1716, 1558-1756},
	url = {https://ieeexplore.ieee.org/document/8809393/},
	doi = {10.1109/MCG.2019.2923483},
	language = {en},
	number = {5},
	urldate = {2026-03-11},
	journal = {IEEE Computer Graphics and Applications},
	author = {Wang, Yun and Segal, Adrien and Klatzky, Roberta and Keefe, Daniel F. and Isenberg, Petra and Hurtienne, Jorn and Hornecker, Eva and Dwyer, Tim and Barrass, Stephen},
	month = sep,
	year = {2019},
	pages = {8--17},
}

@article{Isenberg:2017:VMC,
  author      = {Petra Isenberg and Florian Heimerl and Steffen Koch and Tobias Isenberg and Panpan Xu and Chad Stolper and Michael Sedlmair and Jian Chen and Torsten M{\"o}ller and John Stasko},
  title       = {vispubdata.org: A Metadata Collection about {IEEE} Visualization ({VIS}) Publications},
  journal     = {IEEE Transactions on Visualization and Computer Graphics},
  year        = {2017},
  volume      = {23},
  number      = {9},
  month       = sep,
  pages       = {2199--2206},
  doi         = {10.1109/TVCG.2016.2615308},
  doi_url     = {https://doi.org/10.1109/TVCG.2016.2615308},
  url         = {https://tobias.isenberg.cc/VideosAndDemos/Isenberg2017VMC},
  url2        = {http://www.vispubdata.org/site/vispubdata/},
}

@book{trochim_2006,
  author    = {Trochim, William M. K. and Donnelly, James P.},
  title     = {The Research Methods Knowledge Base},
  edition   = {3rd},
  year      = {2006},
  publisher = {Atomic Dog},
  address   = {Cincinnati, OH}
}

@book{rubin_2023,
  author    = {Rubin, Rick},
  title     = {The Creative Act: A Way of Being},
  year      = {2023},
  publisher = {Penguin Press},
  address   = {New York, NY}
}

@misc{cai_2026,
  title = {Through a Live Elections Dashboard, Darkly: Managing Expectations and Trust in Progressive Vote Counting During the 2024 U.S. Election},
  author = {Cai, Mandi and Wang, Grace and Mortenson, Chloe and Yang, Fumeng and Nisbet, Erik and Kay, Matthew},
  year = {2026},
  doi = {10.17605/OSF.IO/QXK2T},
  url = {https://doi.org/10.17605/OSF.IO/QXK2T},
  note = {OSF Project}
}

@inproceedings{cai_watching_2024,
	address = {Honolulu HI USA},
	title = {Watching the {Election} {Sausage} {Get} {Made}: {How} {Data} {Journalists} {Visualize} the {Vote} {Counting} {Process} in {U}.{S}. {Elections}},
	isbn = {979-8-4007-0330-0},
	shorttitle = {Watching the {Election} {Sausage} {Get} {Made}},
	url = {https://dl.acm.org/doi/10.1145/3613904.3642329},
	doi = {10.1145/3613904.3642329},
	language = {en},
	urldate = {2024-06-05},
	booktitle = {Proceedings of the {CHI} {Conference} on {Human} {Factors} in {Computing} {Systems}},
	publisher = {ACM},
	author = {Cai, Mandi and Kay, Matthew},
	month = may,
	year = {2024},
	pages = {1--19},
}

@book{hawking_2010,
  author    = {Hawking, Stephen and Mlodinow, Leonard},
  title     = {The Grand Design},
  year      = {2010},
  publisher = {Bantam},
  address   = {New York},
}

@incollection{cobley_3_2013,
	title = {3 {Constructing} theories in communication research},
	isbn = {978-3-11-024044-3},
	url = {https://www.degruyter.com/document/doi/10.1515/9783110240450.39/html},
	doi = {10.1515/9783110240450.39},
	language = {en},
	urldate = {2024-04-02},
	booktitle = {Theories and {Models} of {Communication}},
	publisher = {DE GRUYTER},
	author = {Craig, Robert T.},
	editor = {Cobley, Paul and Schulz, Peter J.},
	month = jan,
	year = {2013},
	pages = {39--58},
}

@misc{synecdoche_2008,
  title        = {{Synecdoche, New York}},
  year         = {2008},
  author       = {Charlie Kaufman},
  howpublished = {Film},
  note         = {Sony Pictures Classics}
}

@book{jaccard_2020,
  author    = {Jaccard, James and Jacoby, Jacob},
  title     = {Theory Construction and Model-Building Skills},
  year      = {2020},
  publisher = {The Guilford Press},
  address   = {New York},
}

@misc{allegory_2026,
	title = {Allegory of the cave},
    author = {Wikipedia},
	url = {https://en.wikipedia.org/w/index.php?title=Allegory_of_the_cave&oldid=1343665564},
	language = {en},
	urldate = {2026-03-29},
	month = mar,
	year = {2026},
}

@inproceedings{varona_theory_2025,
	title = {Theory is {Shapes}},
	language = {en},
	urldate = {2026-03-28},
	author = {Varona, Matthew and Hedayati, Maryam and Kay, Matthew and Nobre, Carolina},
    doi={10.48550/arXiv.2510.01382},
    url={https://doi.org/10.48550/arXiv.2510.01382},    
    booktitle = {Proceedings of alt.vis 2025},
	month = oct,
	year = {2025},
}

@article{busse_boundary_2017,
	title = {Boundary {Conditions}: {What} {They} {Are}, {How} to {Explore} {Them}, {Why} {We} {Need} {Them}, and {When} to {Consider} {Them}},
	volume = {20},
	issn = {1094-4281, 1552-7425},
	shorttitle = {Boundary {Conditions}},
	url = {http://journals.sagepub.com/doi/10.1177/1094428116641191},
	doi = {10.1177/1094428116641191},
	language = {en},
	number = {4},
	urldate = {2023-10-03},
	journal = {Organizational Research Methods},
	author = {Busse, Christian and Kach, Andrew P. and Wagner, Stephan M.},
	month = oct,
	year = {2017},
	pages = {574--609},
}

@article{yarkoni_generalizability_2022,
	title = {The generalizability crisis},
	volume = {45},
	issn = {0140-525X, 1469-1825},
	url = {https://www.cambridge.org/core/product/identifier/S0140525X20001685/type/journal_article},
	doi = {10.1017/S0140525X20001685},
	language = {en},
	urldate = {2025-09-17},
	journal = {Behavioral and Brain Sciences},
	author = {Yarkoni, Tal},
	year = {2022},
	pages = {e1},
}

@article{pearl_external_2014,
	title = {External {Validity}: {From} {Do}-{Calculus} to {Transportability} {Across} {Populations}},
	volume = {29},
	issn = {0883-4237},
	shorttitle = {External {Validity}},
	url = {https://projecteuclid.org/journals/statistical-science/volume-29/issue-4/External-Validity-From-Do-Calculus-to-Transportability-Across-Populations/10.1214/14-STS486.full},
	doi = {10.1214/14-STS486},
	language = {en},
	number = {4},
	urldate = {2026-03-16},
	journal = {Statistical Science},
	author = {Pearl, Judea and Bareinboim, Elias},
	month = nov,
	year = {2014},
}

@article{lesko_generalizing_2017,
	title = {Generalizing {Study} {Results}: {A} {Potential} {Outcomes} {Perspective}},
	volume = {28},
	issn = {1044-3983},
	shorttitle = {Generalizing {Study} {Results}},
	url = {http://journals.lww.com/00001648-201707000-00013},
	doi = {10.1097/EDE.0000000000000664},
	language = {en},
	number = {4},
	urldate = {2026-03-16},
	journal = {Epidemiology},
	author = {Lesko, Catherine R. and Buchanan, Ashley L. and Westreich, Daniel and Edwards, Jessie K. and Hudgens, Michael G. and Cole, Stephen R.},
	month = jul,
	year = {2017},
	pages = {553--561},
}

@article{akbaba_entanglements_2025,
	title = {Entanglements for {Visualization}: {Changing} {Research} {Outcomes} through {Feminist} {Theory}},
	volume = {31},
	copyright = {https://ieeexplore.ieee.org/Xplorehelp/downloads/license-information/IEEE.html},
	issn = {1077-2626, 1941-0506, 2160-9306},
	shorttitle = {Entanglements for {Visualization}},
	url = {https://ieeexplore.ieee.org/document/10670254/},
	doi = {10.1109/TVCG.2024.3456171},
	language = {en},
	number = {1},
	urldate = {2025-03-28},
	journal = {IEEE Transactions on Visualization and Computer Graphics},
	author = {Akbaba, Derya and Klein, Lauren and Meyer, Miriah},
	month = jan,
	year = {2025},
	pages = {1279--1289},
}

@incollection{hutchison_creation_2008,
  title={Creation and collaboration: Engaging new audiences for information visualization},
  author={Heer, Jeffrey and Van Ham, Frank and Carpendale, Sheelagh and Weaver, Chris and Isenberg, Petra},
  booktitle={Information Visualization: Human-Centered Issues and Perspectives},
  pages={92--133},
  year={2008},
  publisher={Springer}
}

@phdthesis{danziger_information_2008,
  title={Information visualization for the people},
  author={Danziger, Michael J},
  year={2008},
  school={Massachusetts Institute of Technology, Department of Comparative Media Studies}
}

@article{fu_more_2024,
	title = {More {Than} {Data} {Stories}: {Broadening} the {Role} of {Visualization} in {Contemporary} {Journalism}},
	volume = {30},
	copyright = {https://ieeexplore.ieee.org/Xplorehelp/downloads/license-information/IEEE.html},
	issn = {1077-2626, 1941-0506, 2160-9306},
	shorttitle = {More {Than} {Data} {Stories}},
	url = {https://ieeexplore.ieee.org/document/10158423/},
	doi = {10.1109/TVCG.2023.3287585},
	language = {en},
	number = {8},
	urldate = {2025-08-16},
	journal = {IEEE Transactions on Visualization and Computer Graphics},
	author = {Fu, Yu and Stasko, John},
	month = aug,
	year = {2024},
	pages = {5240--5259},
}

@article{lee_reaching_2020,
	title = {Reaching {Broader} {Audiences} {With} {Data} {Visualization}},
	volume = {40},
	copyright = {https://ieeexplore.ieee.org/Xplorehelp/downloads/license-information/IEEE.html},
	issn = {0272-1716, 1558-1756},
	url = {https://ieeexplore.ieee.org/document/9023497/},
	doi = {10.1109/MCG.2020.2968244},
	language = {en},
	number = {2},
	urldate = {2025-11-03},
	journal = {IEEE Computer Graphics and Applications},
	author = {Lee, Bongshin and Choe, Eun Kyoung and Isenberg, Petra and Marriott, Kim and Stasko, John},
	month = mar,
	year = {2020},
	pages = {82--90},
}

@article{pousman_casual_2007,
	title = {Casual {Information} {Visualization}: {Depictions} of {Data} in {Everyday} {Life}},
	volume = {13},
	copyright = {https://ieeexplore.ieee.org/Xplorehelp/downloads/license-information/IEEE.html},
	issn = {1077-2626},
	shorttitle = {Casual {Information} {Visualization}},
	url = {http://ieeexplore.ieee.org/document/4376134/},
	doi = {10.1109/TVCG.2007.70541},
	language = {en},
	number = {6},
	urldate = {2026-03-11},
	journal = {IEEE Transactions on Visualization and Computer Graphics},
	author = {Pousman, Zachary and Stasko, John and Mateas, Michael},
	month = nov,
	year = {2007},
	pages = {1145--1152},
}

@article{park_positivism_2020,
	title = {The {Positivism} {Paradigm} of {Research}},
	volume = {95},
	issn = {1040-2446},
	url = {https://journals.lww.com/10.1097/ACM.0000000000003093},
	doi = {10.1097/ACM.0000000000003093},
	language = {en},
	number = {5},
	urldate = {2026-03-13},
	journal = {Academic Medicine},
	author = {Park, Yoon Soo and Konge, Lars and Artino, Anthony R.},
	month = may,
	year = {2020},
	pages = {690--694},
}

@article{wall_vishikers_2022,
	title = {{VisHikers}’ {Guide} to {Evaluation}: {Competing} {Considerations} in {Study} {Design}},
	volume = {42},
	copyright = {https://ieeexplore.ieee.org/Xplorehelp/downloads/license-information/IEEE.html},
	issn = {0272-1716, 1558-1756},
	shorttitle = {{VisHikers}’ {Guide} to {Evaluation}},
	url = {https://ieeexplore.ieee.org/document/9790020/},
	doi = {10.1109/MCG.2022.3152676},
	language = {en},
	number = {3},
	urldate = {2026-03-18},
	journal = {IEEE Computer Graphics and Applications},
	author = {Wall, Emily and Xiong, Cindy and Kim, Yea-Seul},
	month = may,
	year = {2022},
	pages = {29--38},
}

@article{ponterotto_qualitative_2005,
	title = {Qualitative research in counseling psychology: {A} primer on research paradigms and philosophy of science.},
	volume = {52},
	issn = {1939-2168, 0022-0167},
	shorttitle = {Qualitative research in counseling psychology},
	url = {https://doi.apa.org/doi/10.1037/0022-0167.52.2.126},
	doi = {10.1037/0022-0167.52.2.126},
	language = {en},
	number = {2},
	urldate = {2026-03-13},
	journal = {Journal of Counseling Psychology},
	author = {Ponterotto, Joseph G.},
	month = apr,
	year = {2005},
	pages = {126--136},
}

@article{tracy_qualitative_2010,
	title = {Qualitative {Quality}: {Eight} “{Big}-{Tent}” {Criteria} for {Excellent} {Qualitative} {Research}},
	volume = {16},
	copyright = {https://journals.sagepub.com/page/policies/text-and-data-mining-license},
	issn = {1077-8004, 1552-7565},
	shorttitle = {Qualitative {Quality}},
	url = {https://journals.sagepub.com/doi/10.1177/1077800410383121},
	doi = {10.1177/1077800410383121},
	language = {en},
	number = {10},
	urldate = {2025-10-27},
	journal = {Qualitative Inquiry},
	author = {Tracy, Sarah J.},
	month = dec,
	year = {2010},
	pages = {837--851},
}

@article{meyer_criteria_2019,
	title = {Criteria for {Rigor} in {Visualization} {Design} {Study}},
	copyright = {https://ieeexplore.ieee.org/Xplorehelp/downloads/license-information/IEEE.html},
	issn = {1077-2626, 1941-0506, 2160-9306},
	url = {https://ieeexplore.ieee.org/document/8809711/},
	doi = {10.1109/TVCG.2019.2934539},
	language = {en},
	urldate = {2025-04-17},
	journal = {IEEE Transactions on Visualization and Computer Graphics},
	author = {Meyer, Miriah and Dykes, Jason},
	year = {2019},
	pages = {1--1},
}

@article{munzner_nested_2009,
	title = {A {Nested} {Model} for {Visualization} {Design} and {Validation}},
	volume = {15},
	copyright = {https://ieeexplore.ieee.org/Xplorehelp/downloads/license-information/IEEE.html},
	issn = {1077-2626},
	url = {http://ieeexplore.ieee.org/document/5290695/},
	doi = {10.1109/TVCG.2009.111},
	language = {en},
	number = {6},
	urldate = {2026-01-25},
	journal = {IEEE Transactions on Visualization and Computer Graphics},
	author = {Munzner, Tamara},
	month = nov,
	year = {2009},
	pages = {921--928},
}

@article{sedlmair_design_2012,
	title = {Design {Study} {Methodology}: {Reflections} from the {Trenches} and the {Stacks}},
	volume = {18},
	copyright = {https://ieeexplore.ieee.org/Xplorehelp/downloads/license-information/IEEE.html},
	issn = {1077-2626},
	shorttitle = {Design {Study} {Methodology}},
	url = {http://ieeexplore.ieee.org/document/6327248/},
	doi = {10.1109/TVCG.2012.213},
	language = {en},
	number = {12},
	urldate = {2026-03-11},
	journal = {IEEE Transactions on Visualization and Computer Graphics},
	author = {Sedlmair, Michael and Meyer, Miriah and Munzner, Tamara},
	month = dec,
	year = {2012},
	pages = {2431--2440},
}

@article{lam_empirical_2012,
	title = {Empirical {Studies} in {Information} {Visualization}: {Seven} {Scenarios}},
	volume = {18},
	copyright = {https://ieeexplore.ieee.org/Xplorehelp/downloads/license-information/IEEE.html},
	issn = {1077-2626},
	shorttitle = {Empirical {Studies} in {Information} {Visualization}},
	url = {http://ieeexplore.ieee.org/document/6095544/},
	doi = {10.1109/TVCG.2011.279},
	language = {en},
	number = {9},
	urldate = {2026-02-05},
	journal = {IEEE Transactions on Visualization and Computer Graphics},
	author = {Lam, H. and Bertini, E. and Isenberg, P. and Plaisant, C. and Carpendale, S.},
	month = sep,
	year = {2012},
	pages = {1520--1536},
}

@article{isenberg2013systematic,
  title={A systematic review on the practice of evaluating visualization},
  author={Isenberg, Tobias and Isenberg, Petra and Chen, Jian and Sedlmair, Michael and M{\"o}ller, Torsten},
  journal={IEEE Transactions on Visualization and Computer Graphics},
  volume={19},
  number={12},
  pages={2818--2827},
  year={2013},
  publisher={IEEE}
}

@inproceedings{twidale_situated_1994,
	address = {Chapel Hill, North Carolina, United States},
	title = {Situated evaluation for cooperative systems},
	copyright = {https://www.acm.org/publications/policies/copyright\_policy\#Background},
	isbn = {978-0-89791-689-9},
	url = {http://portal.acm.org/citation.cfm?doid=192844.193066},
	doi = {10.1145/192844.193066},
	language = {en},
	urldate = {2026-03-18},
	booktitle = {Proceedings of the 1994 {ACM} conference on {Computer} supported cooperative work  - {CSCW} '94},
	publisher = {ACM Press},
	author = {Twidale, Michael and Randall, David and Bentley, Richard},
	year = {1994},
	pages = {441--452},
}

@incollection{kerren_evaluating_2008,
	address = {Berlin, Heidelberg},
	title = {Evaluating {Information} {Visualizations}},
	volume = {4950},
	isbn = {978-3-540-70955-8 978-3-540-70956-5},
	issn = {0302-9743, 1611-3349},
	url = {http://link.springer.com/10.1007/978-3-540-70956-5_2},
	doi = {10.1007/978-3-540-70956-5_2},
	language = {en},
	urldate = {2025-09-18},
	booktitle = {Information {Visualization}},
	publisher = {Springer Berlin Heidelberg},
	author = {Carpendale, Sheelagh},
	editor = {Kerren, Andreas and Stasko, John T. and Fekete, Jean-Daniel and North, Chris},
	year = {2008},
	note = {Series Title: Lecture Notes in Computer Science},
	pages = {19--45},
}

@inproceedings{andrews_evaluation_2008,
  title={Evaluation comes in many guises},
  author={Andrews, Keith},
  booktitle={AVI Workshop on BEyond time and errors (BELIV) Position Paper},
  pages={7--8},
  year={2008}
}

@inproceedings{brown_into_2011,
	address = {Vancouver BC Canada},
	title = {Into the wild: challenges and opportunities for field trial methods},
	isbn = {978-1-4503-0228-9},
	shorttitle = {Into the wild},
	url = {https://dl.acm.org/doi/10.1145/1978942.1979185},
	doi = {10.1145/1978942.1979185},
	language = {en},
	urldate = {2026-02-05},
	booktitle = {Proceedings of the {SIGCHI} {Conference} on {Human} {Factors} in {Computing} {Systems}},
	publisher = {ACM},
	author = {Brown, Barry and Reeves, Stuart and Sherwood, Scott},
	month = may,
	year = {2011},
	pages = {1657--1666},
}

@article{kosara_thoughts_2003,
  title={User studies: Why, how, and when?},
  author={Kosara, Robert and Healey, Christopher G and Interrante, Victoria and Laidlaw, David H and Ware, Colin},
  journal={IEEE Computer Graphics and Applications},
  volume={23},
  number={04},
  pages={20--25},
  year={2003},
  publisher={IEEE Computer Society}
}

@inproceedings{zuk_heuristics_2006,
	address = {Venice Italy},
	title = {Heuristics for information visualization evaluation},
	isbn = {978-1-59593-562-5},
	url = {https://dl.acm.org/doi/10.1145/1168149.1168162},
	doi = {10.1145/1168149.1168162},
	language = {en},
	urldate = {2026-02-05},
	booktitle = {Proceedings of the 2006 {AVI} workshop on {BEyond} time and errors: novel evaluation methods for information visualization},
	publisher = {ACM},
	author = {Zuk, Torre and Schlesier, Lothar and Neumann, Petra and Hancock, Mark S. and Carpendale, Sheelagh},
	month = may,
	year = {2006},
	pages = {1--6},
}

@inproceedings{faisal_internalization_2008,
	address = {Florence Italy},
	title = {Internalization, qualitative methods, and evaluation},
	isbn = {978-1-60558-016-6},
	url = {https://dl.acm.org/doi/10.1145/1377966.1377973},
	doi = {10.1145/1377966.1377973},
	language = {en},
	urldate = {2026-02-05},
	booktitle = {Proceedings of the 2008 {Workshop} on {BEyond} time and errors: novel {evaLuation} methods for {Information} {Visualization}},
	publisher = {ACM},
	author = {Faisal, Sarah and Craft, Brock and Cairns, Paul and Blandford, Ann},
	month = apr,
	year = {2008},
	pages = {1--8},
}

@inproceedings{shneiderman_strategies_2006,
	address = {Venice Italy},
	title = {Strategies for evaluating information visualization tools: multi-dimensional in-depth long-term case studies},
	isbn = {978-1-59593-562-5},
	shorttitle = {Strategies for evaluating information visualization tools},
	url = {https://dl.acm.org/doi/10.1145/1168149.1168158},
	doi = {10.1145/1168149.1168158},
	language = {en},
	urldate = {2026-02-05},
	booktitle = {Proceedings of the 2006 {AVI} workshop on {BEyond} time and errors: novel evaluation methods for information visualization},
	publisher = {ACM},
	author = {Shneiderman, Ben and Plaisant, Catherine},
	month = may,
	year = {2006},
	pages = {1--7},
}

@article{saraiya_insight-based_2005,
	title = {An {Insight}-{Based} {Methodology} for {Evaluating} {Bioinformatics} {Visualizations}},
	volume = {11},
	copyright = {https://ieeexplore.ieee.org/Xplorehelp/downloads/license-information/IEEE.html},
	issn = {1077-2626},
	url = {http://ieeexplore.ieee.org/document/1432690/},
	doi = {10.1109/TVCG.2005.53},
	language = {en},
	number = {4},
	urldate = {2026-03-11},
	journal = {IEEE Transactions on Visualization and Computer Graphics},
	author = {Saraiya, P. and North, C. and Duca, K.},
	month = jul,
	year = {2005},
	pages = {443--456},
}

@article{saraiya_insight-based_2006,
	title = {An {Insight}-{Based} {Longitudinal} {Study} of {Visual} {Analytics}},
	volume = {12},
	copyright = {https://ieeexplore.ieee.org/Xplorehelp/downloads/license-information/IEEE.html},
	issn = {1077-2626},
	url = {http://ieeexplore.ieee.org/document/1703371/},
	doi = {10.1109/TVCG.2006.85},
	language = {en},
	number = {6},
	urldate = {2026-03-11},
	journal = {IEEE Transactions on Visualization and Computer Graphics},
	author = {Saraiya, P. and North, C. and {Vy Lam} and Duca, K.A.},
	month = nov,
	year = {2006},
	pages = {1511--1522},
}

@article{mckenna_design_2014,
	title = {Design {Activity} {Framework} for {Visualization} {Design}},
	volume = {20},
	copyright = {https://ieeexplore.ieee.org/Xplorehelp/downloads/license-information/IEEE.html},
	issn = {1077-2626},
	url = {http://ieeexplore.ieee.org/document/6876000/},
	doi = {10.1109/TVCG.2014.2346331},
	language = {en},
	number = {12},
	urldate = {2026-02-05},
	journal = {IEEE Transactions on Visualization and Computer Graphics},
	author = {McKenna, Sean and Mazur, Dominika and Agutter, James and Meyer, Miriah},
	month = dec,
	year = {2014},
	pages = {2191--2200},
}

@inproceedings{isenberg_grounded_2008,
  title={Grounded evaluation of information visualizations},
  author={Isenberg, Petra and Zuk, Torre and Collins, Christopher and Carpendale, Sheelagh},
  booktitle={Proceedings of the 2008 Workshop on BEyond time and errors: novel evaLuation methods for Information Visualization},
  pages={1--8},
  year={2008}
}

@article{yang_swaying_2023,
	title = {Swaying the {Public}? {Impacts} of {Election} {Forecast} {Visualizations} on {Emotion}, {Trust}, and {Intention} in the 2022 {U}.{S}. {Midterms}},
	copyright = {https://ieeexplore.ieee.org/Xplorehelp/downloads/license-information/IEEE.html},
	issn = {1077-2626, 1941-0506, 2160-9306},
	shorttitle = {Swaying the {Public}?},
	url = {https://ieeexplore.ieee.org/document/10309864/},
	doi = {10.1109/TVCG.2023.3327356},
	language = {en},
	urldate = {2026-01-07},
	journal = {IEEE Transactions on Visualization and Computer Graphics},
	author = {Yang, Fumeng and Cai, Mandi and Mortenson, Chloe and Fakhari, Hoda and Lokmanoglu, Ayse D. and Hullman, Jessica and Franconeri, Steven and Diakopoulos, Nicholas and Nisbet, Erik C. and Kay, Matthew},
	year = {2023},
	pages = {1--11},
}

@inproceedings{schuster_who_2023,
  title={Practitioners' Perspectives on Designing Data Visualizations for the General Public},
  author={Schuster, Regina and Gregory, Kathleen and M{\"o}ller, Torsten and Koesten, Laura},
  booktitle={Proceedings of the 2026 CHI Conference on Human Factors in Computing Systems},
  pages={1--19},
  year={2026}
}

@inproceedings{arunkumar_lost_2025,
	address = {Yokohama Japan},
	title = {Lost in {Translation}: {How} {Does} {Bilingualism} {Shape} {Reader} {Preferences} for {Annotated} {Charts}?},
	isbn = {979-8-4007-1394-1},
	shorttitle = {Lost in {Translation}},
	url = {https://dl.acm.org/doi/10.1145/3706598.3713380},
	doi = {10.1145/3706598.3713380},
	language = {en},
	urldate = {2026-02-19},
	booktitle = {Proceedings of the 2025 {CHI} {Conference} on {Human} {Factors} in {Computing} {Systems}},
	publisher = {ACM},
	author = {Arunkumar, Anjana and Padilla, Lace M. and Bryan, Chris},
	month = apr,
	year = {2025},
	pages = {1--22},
}

@inproceedings{peck_data_2019,
	address = {Glasgow Scotland Uk},
	title = {Data is {Personal}: {Attitudes} and {Perceptions} of {Data} {Visualization} in {Rural} {Pennsylvania}},
	isbn = {978-1-4503-5970-2},
	shorttitle = {Data is {Personal}},
	url = {https://dl.acm.org/doi/10.1145/3290605.3300474},
	doi = {10.1145/3290605.3300474},
	language = {en},
	urldate = {2023-01-17},
	booktitle = {Proceedings of the 2019 {CHI} {Conference} on {Human} {Factors} in {Computing} {Systems}},
	publisher = {ACM},
	author = {Peck, Evan M. and Ayuso, Sofia E. and El-Etr, Omar},
	month = may,
	year = {2019},
	pages = {1--12},
}

@article{dasu_sea_2021,
	title = {Sea of {Genes}: {A} {Reflection} on {Visualising} {Metagenomic} {Data} for {Museums}},
	volume = {27},
	copyright = {https://ieeexplore.ieee.org/Xplorehelp/downloads/license-information/IEEE.html},
	issn = {1077-2626, 1941-0506, 2160-9306},
	shorttitle = {Sea of {Genes}},
	url = {https://ieeexplore.ieee.org/document/9241413/},
	doi = {10.1109/TVCG.2020.3030412},
	language = {en},
	number = {2},
	urldate = {2026-01-07},
	journal = {IEEE Transactions on Visualization and Computer Graphics},
	author = {Dasu, Keshav and Ma, Kwan-Liu and Ma, Joyce and Frazier, Jennifer},
	month = feb,
	year = {2021},
	pages = {935--945},
}

@article{padilla_multiple_2022,
	title = {Multiple {Forecast} {Visualizations} ({MFVs}): {Trade}-offs in {Trust} and {Performance} in {Multiple} {COVID}-19 {Forecast} {Visualizations}},
	copyright = {https://creativecommons.org/licenses/by/4.0/legalcode},
	issn = {1077-2626, 1941-0506, 2160-9306},
	shorttitle = {Multiple {Forecast} {Visualizations} ({MFVs})},
	url = {https://ieeexplore.ieee.org/document/9904455/},
	doi = {10.1109/TVCG.2022.3209457},
	language = {en},
	urldate = {2026-01-06},
	journal = {IEEE Transactions on Visualization and Computer Graphics},
	author = {Padilla, Lace and Fygenson, Racquel and Castro, Spencer C. and Bertini, Enrico},
	year = {2022},
	pages = {1--11},
}

@article{sterzik_perception_2023,
	title = {Perception of {Line} {Attributes} for {Visualization}},
	copyright = {https://ieeexplore.ieee.org/Xplorehelp/downloads/license-information/IEEE.html},
	issn = {1077-2626, 1941-0506, 2160-9306},
	url = {https://ieeexplore.ieee.org/document/10292709/},
	doi = {10.1109/TVCG.2023.3326523},
	language = {en},
	urldate = {2026-01-07},
	journal = {IEEE Transactions on Visualization and Computer Graphics},
	author = {Sterzik, Anna and Lichtenberg, Nils and Wilms, Jana and Krone, Michael and Cunningham, Douglas W. and Lawonn, Kai},
	year = {2023},
	pages = {1--11},
}

@inproceedings{panavas_juvenile_2022,
	address = {New Orleans LA USA},
	title = {Juvenile {Graphical} {Perception}: {A} {Comparison} between {Children} and {Adults}},
	copyright = {https://www.acm.org/publications/policies/copyright\_policy\#Background},
	isbn = {978-1-4503-9157-3},
	shorttitle = {Juvenile {Graphical} {Perception}},
	url = {https://dl.acm.org/doi/10.1145/3491102.3501893},
	doi = {10.1145/3491102.3501893},
	language = {en},
	urldate = {2026-01-04},
	booktitle = {{CHI} {Conference} on {Human} {Factors} in {Computing} {Systems}},
	publisher = {ACM},
	author = {Panavas, Liudas and Worth, Amy E. and Crnovrsanin, Tarik and Sathyamurthi, Tejas and Cordes, Sara and Borkin, Michelle A. and Dunne, Cody},
	month = apr,
	year = {2022},
	pages = {1--14},
}

@article{stokes_striking_2022,
	title = {Striking a {Balance}: {Reader} {Takeaways} and {Preferences} when {Integrating} {Text} and {Charts}},
	issn = {1077-2626, 1941-0506, 2160-9306},
	shorttitle = {Striking a {Balance}},
	url = {https://ieeexplore.ieee.org/document/9904452/},
	doi = {10.1109/TVCG.2022.3209383},
	urldate = {2022-11-02},
	journal = {IEEE Transactions on Visualization and Computer Graphics},
	author = {Stokes, Chase and Setlur, Vidya and Cogley, Bridget and Satyanarayan, Arvind and Hearst, Marti A.},
	year = {2022},
	pages = {1--11},
}

@article{kauer_discursive_2025,
	title = {Discursive {Patinas}: {Anchoring} {Discussions} in {Data} {Visualizations}},
	volume = {31},
	copyright = {https://ieeexplore.ieee.org/Xplorehelp/downloads/license-information/IEEE.html},
	issn = {1077-2626, 1941-0506, 2160-9306},
	shorttitle = {Discursive {Patinas}},
	url = {https://ieeexplore.ieee.org/document/10679900/},
	doi = {10.1109/TVCG.2024.3456334},
	language = {en},
	number = {1},
	urldate = {2026-03-09},
	journal = {IEEE Transactions on Visualization and Computer Graphics},
	author = {Kauer, Tobias and Akbaba, Derya and Dörk, Marian and Bach, Benjamin},
	month = jan,
	year = {2025},
	pages = {1246--1256},
}

@inproceedings{akbaba_troubling_2023,
	address = {Hamburg Germany},
	title = {Troubling {Collaboration}: {Matters} of {Care} for {Visualization} {Design} {Study}},
	isbn = {978-1-4503-9421-5},
	shorttitle = {Troubling {Collaboration}},
	url = {https://dl.acm.org/doi/10.1145/3544548.3581168},
	doi = {10.1145/3544548.3581168},
	language = {en},
	urldate = {2023-05-13},
	booktitle = {Proceedings of the 2023 {CHI} {Conference} on {Human} {Factors} in {Computing} {Systems}},
	publisher = {ACM},
	author = {Akbaba, Derya and Lange, Devin and Correll, Michael and Lex, Alexander and Meyer, Miriah},
	month = apr,
	year = {2023},
	pages = {1--15},
}

@article{ma_decoding_2019,
	title = {Decoding a {Complex} {Visualization} in a {Science} {Museum} - {An} {Empirical} {Study}},
	copyright = {https://ieeexplore.ieee.org/Xplorehelp/downloads/license-information/IEEE.html},
	issn = {1077-2626, 1941-0506, 2160-9306},
	url = {https://ieeexplore.ieee.org/document/8805458/},
	doi = {10.1109/TVCG.2019.2934401},
	language = {en},
	urldate = {2025-11-03},
	journal = {IEEE Transactions on Visualization and Computer Graphics},
	author = {Ma, Joyce and Ma, Kwan-Liu and Frazier, Jennifer},
	year = {2019},
	pages = {1--1},
}

@inproceedings{amar_low-level_2005,
	address = {Minneapolis, MN, USA},
	title = {Low-level components of analytic activity in information visualization},
	isbn = {978-0-7803-9464-3},
	url = {http://ieeexplore.ieee.org/document/1532136/},
	doi = {10.1109/INFVIS.2005.1532136},
	language = {en},
	urldate = {2026-02-10},
	booktitle = {{IEEE} {Symposium} on {Information} {Visualization}, 2005. {INFOVIS} 2005.},
	publisher = {IEEE},
	author = {Amar, R. and Eagan, J. and Stasko, J.},
	year = {2005},
	pages = {111--117},
}

@inproceedings{burns_how_2020,
  title={How to evaluate data visualizations across different levels of understanding},
  author={Burns, Alyxander and Xiong, Cindy and Franconeri, Steven and Cairo, Alberto and Mahyar, Narges},
  booktitle={2020 IEEE Workshop on Evaluation and Beyond-Methodological Approaches to Visualization (BELIV)},
  pages={19--28},
  year={2020},
  organization={IEEE}
}

@inproceedings{lee_task_2006,
	address = {Venice Italy},
	title = {Task taxonomy for graph visualization},
	isbn = {978-1-59593-562-5},
	url = {https://dl.acm.org/doi/10.1145/1168149.1168168},
	doi = {10.1145/1168149.1168168},
	language = {en},
	urldate = {2026-02-10},
	booktitle = {Proceedings of the 2006 {AVI} workshop on {BEyond} time and errors: novel evaluation methods for information visualization},
	publisher = {ACM},
	author = {Lee, Bongshin and Plaisant, Catherine and Parr, Cynthia Sims and Fekete, Jean-Daniel and Henry, Nathalie},
	month = may,
	year = {2006},
	pages = {1--5},
}

@article{sarikaya_scatterplots_2018,
	title = {Scatterplots: {Tasks}, {Data}, and {Designs}},
	volume = {24},
	issn = {1077-2626},
	shorttitle = {Scatterplots},
	url = {http://ieeexplore.ieee.org/document/8017602/},
	doi = {10.1109/TVCG.2017.2744184},
	language = {en},
	number = {1},
	urldate = {2023-02-10},
	journal = {IEEE Transactions on Visualization and Computer Graphics},
	author = {Sarikaya, Alper and Gleicher, Michael},
	month = jan,
	year = {2018},
	pages = {402--412},
}

@article{lan_kineticharts_2022,
	title = {Kineticharts: {Augmenting} {Affective} {Expressiveness} of {Charts} in {Data} {Stories} with {Animation} {Design}},
	volume = {28},
	copyright = {https://ieeexplore.ieee.org/Xplorehelp/downloads/license-information/IEEE.html},
	issn = {1077-2626, 1941-0506, 2160-9306},
	shorttitle = {Kineticharts},
	url = {https://ieeexplore.ieee.org/document/9552240/},
	doi = {10.1109/TVCG.2021.3114775},
	language = {en},
	number = {1},
	urldate = {2026-01-05},
	journal = {IEEE Transactions on Visualization and Computer Graphics},
	author = {Lan, Xingyu and Shi, Yang and Wu, Yanqiu and Jiao, Xiaohan and Cao, Nan},
	month = jan,
	year = {2022},
	pages = {933--943},
}

@article{morais_exploring_2022,
	title = {Exploring {How} {Visualization} {Design} and {Situatedness} {Evoke} {Compassion} in the {Wild}},
	volume = {41},
	issn = {0167-7055, 1467-8659},
	url = {https://onlinelibrary.wiley.com/doi/10.1111/cgf.14553},
	doi = {10.1111/cgf.14553},
	language = {en},
	number = {3},
	urldate = {2026-01-04},
	journal = {Computer Graphics Forum},
	author = {Morais, L. and Andrade, N. and Sousa, D.},
	month = jun,
	year = {2022},
	pages = {441--452},
}

@article{liem_structure_2020,
	title = {Structure and {Empathy} in {Visual} {Data} {Storytelling}: {Evaluating} their {Influence} on {Attitude}},
	volume = {39},
	issn = {0167-7055, 1467-8659},
	shorttitle = {Structure and {Empathy} in {Visual} {Data} {Storytelling}},
	url = {https://onlinelibrary.wiley.com/doi/10.1111/cgf.13980},
	doi = {10.1111/cgf.13980},
	language = {en},
	number = {3},
	urldate = {2026-01-04},
	journal = {Computer Graphics Forum},
	author = {Liem, J. and Perin, C. and Wood, J.},
	month = jun,
	year = {2020},
	pages = {277--289},
}

@article{sprague_exploring_2012,
	title = {Exploring how and why people use visualizations in casual contexts: {Modeling} user goals and regulated motivations},
	volume = {11},
	issn = {1473-8716, 1473-8724},
	shorttitle = {Exploring how and why people use visualizations in casual contexts},
	url = {https://journals.sagepub.com/doi/10.1177/1473871611433710},
	doi = {10.1177/1473871611433710},
	language = {en},
	number = {2},
	urldate = {2025-11-03},
	journal = {Information Visualization},
	author = {Sprague, David and Tory, Melanie},
	month = apr,
	year = {2012},
	pages = {106--123},
}

@article{kihlstrom_ecological_2021,
	title = {Ecological {Validity} and “{Ecological} {Validity}”},
	volume = {16},
	copyright = {https://journals.sagepub.com/page/policies/text-and-data-mining-license},
	issn = {1745-6916, 1745-6924},
	url = {https://journals.sagepub.com/doi/10.1177/1745691620966791},
	doi = {10.1177/1745691620966791},
	language = {en},
	number = {2},
	urldate = {2025-07-21},
	journal = {Perspectives on Psychological Science},
	publisher = {SAGE Publications},
	author = {Kihlstrom, John F.},
	month = mar,
	year = {2021},
	pages = {466--471},
}

@article{lu_ecological_2025,
	title = {Ecological validity of {VR} experiments for psychological and physiological responses to audio-visual environments},
	volume = {274},
	issn = {03601323},
	url = {https://linkinghub.elsevier.com/retrieve/pii/S0360132325002495},
	doi = {10.1016/j.buildenv.2025.112767},
	language = {en},
	urldate = {2025-09-17},
	journal = {Building and Environment},
	author = {Lu, Yichun and Lau, Siu-Kit},
	month = apr,
	year = {2025},
	pages = {112767},
}

@article{orne_social_1962,
  author  = {Orne, Martin T.},
  title   = {On the Social Psychology of the Psychological Experiment: With Particular Reference to Demand Characteristics and Their Implications},
  journal = {American Psychologist},
  year    = {1962},
  volume  = {17},
  pages   = {776--783}
}

@inproceedings{ogbonnaya-ogburu_critical_2020,
	address = {Honolulu HI USA},
	title = {Critical {Race} {Theory} for {HCI}},
	isbn = {978-1-4503-6708-0},
	url = {https://dl.acm.org/doi/10.1145/3313831.3376392},
	doi = {10.1145/3313831.3376392},
	language = {en},
	urldate = {2023-04-06},
	booktitle = {Proceedings of the 2020 {CHI} {Conference} on {Human} {Factors} in {Computing} {Systems}},
	publisher = {ACM},
	author = {Ogbonnaya-Ogburu, Ihudiya Finda and Smith, Angela D.R. and To, Alexandra and Toyama, Kentaro},
	month = apr,
	year = {2020},
	pages = {1--16},
}

@article{liang_embracing_2021,
	title = {Embracing {Four} {Tensions} in {Human}-{Computer} {Interaction} {Research} with {Marginalized} {People}},
	volume = {28},
	issn = {1073-0516, 1557-7325},
	url = {https://dl.acm.org/doi/10.1145/3443686},
	doi = {10.1145/3443686},
	language = {en},
	number = {2},
	urldate = {2025-07-29},
	journal = {ACM Transactions on Computer-Human Interaction},
	author = {Liang, Calvin A. and Munson, Sean A. and Kientz, Julie A.},
	month = apr,
	year = {2021},
	pages = {1--47},
}

@book{akbaba_shifting_2025,
	address = {Linköping},
	series = {Linköping {Studies} in {Science} and {Technology}. {Dissertations}},
	title = {Shifting {Perspectives} : {Conducting} {Visualization} {Research} with {Entanglement} {Epistemology}},
	volume = {2459},
	isbn = {978-91-8118-164-7},
	shorttitle = {Shifting {Perspectives}},
	url = {https://urn.kb.se/resolve?urn=urn:nbn:se:liu:diva-214083},
	doi = {10.3384/9789181181654},
	language = {en},
	urldate = {2025-10-10},
	publisher = {Linköping University Electronic Press},
	author = {Akbaba, Derya},
	month = sep,
	year = {2025},
}

@article{meyer_nested_2015,
	title = {The nested blocks and guidelines model},
	volume = {14},
	issn = {1473-8716, 1473-8724},
	url = {https://journals.sagepub.com/doi/10.1177/1473871613510429},
	doi = {10.1177/1473871613510429},
	language = {en},
	number = {3},
	urldate = {2026-02-06},
	journal = {Information Visualization},
	author = {Meyer, Miriah and Sedlmair, Michael and Quinan, P Samuel and Munzner, Tamara},
	month = jul,
	year = {2015},
	pages = {234--249},
}

@inproceedings{plaisant_challenge_2004,
	address = {Gallipoli Italy},
	title = {The challenge of information visualization evaluation},
	isbn = {978-1-58113-867-2},
	url = {https://dl.acm.org/doi/10.1145/989863.989880},
	doi = {10.1145/989863.989880},
	language = {en},
	urldate = {2026-01-25},
	booktitle = {Proceedings of the working conference on {Advanced} visual interfaces},
	publisher = {ACM},
	author = {Plaisant, Catherine},
	month = may,
	year = {2004},
	pages = {109--116},
}

@article{van_laar_relation_2017,
	title = {The relation between 21st-century skills and digital skills: {A} systematic literature review},
	volume = {72},
	issn = {07475632},
	shorttitle = {The relation between 21st-century skills and digital skills},
	url = {https://linkinghub.elsevier.com/retrieve/pii/S0747563217301590},
	doi = {10.1016/j.chb.2017.03.010},
	language = {en},
	urldate = {2025-10-02},
	journal = {Computers in Human Behavior},
	author = {Van Laar, Ester and Van Deursen, Alexander J.A.M. and Van Dijk, Jan A.G.M. and De Haan, Jos},
	month = jul,
	year = {2017},
	pages = {577--588},
}

@article{moher_reprintpreferred_2009,
	title = {Reprint—{Preferred} {Reporting} {Items} for {Systematic} {Reviews} and {Meta}-{Analyses}: {The} {PRISMA} {Statement}},
	volume = {89},
	issn = {0031-9023, 1538-6724},
	shorttitle = {Reprint—{Preferred} {Reporting} {Items} for {Systematic} {Reviews} and {Meta}-{Analyses}},
	url = {https://academic.oup.com/ptj/article/89/9/873/2737590},
	doi = {10.1093/ptj/89.9.873},
	language = {en},
	number = {9},
	urldate = {2025-11-03},
	journal = {Physical Therapy},
	author = {Moher, David and Liberati, Alessandro and Tetzlaff, Jennifer and Altman, Douglas G. and {and the PRISMA Group}},
	month = sep,
	year = {2009},
	pages = {873--880},
}

@misc{data_owid_2024,
	title = {{OWID} {Homepage}},
	url = {https://ourworldindata.org},
	language = {en},
	urldate = {2025-11-03},
	author = {{Our World in Data}},
	year = {2025},
}

@article{hullman_pursuit_2019,
	title = {In {Pursuit} of {Error}: {A} {Survey} of {Uncertainty} {Visualization} {Evaluation}},
	volume = {25},
	issn = {1077-2626, 1941-0506, 2160-9306},
	shorttitle = {In {Pursuit} of {Error}},
	url = {https://ieeexplore.ieee.org/document/8457476/},
	doi = {10.1109/TVCG.2018.2864889},
	number = {1},
	urldate = {2022-11-01},
	journal = {IEEE Transactions on Visualization and Computer Graphics},
	author = {Hullman, Jessica and Qiao, Xiaoli and Correll, Michael and Kale, Alex and Kay, Matthew},
	month = jan,
	year = {2019},
	pages = {903--913},
}

@article{braun_reflecting_2019,
	title = {Reflecting on reflexive thematic analysis},
	volume = {11},
	issn = {2159-676X, 2159-6778},
	url = {https://www.tandfonline.com/doi/full/10.1080/2159676X.2019.1628806},
	doi = {10.1080/2159676X.2019.1628806},
	language = {en},
	number = {4},
	urldate = {2023-11-18},
	journal = {Qualitative Research in Sport, Exercise and Health},
	author = {Braun, Virginia and Clarke, Victoria},
	month = aug,
	year = {2019},
	pages = {589--597},
}

@article{braun_using_2006,
	title = {Using thematic analysis in psychology},
	volume = {3},
	issn = {1478-0887, 1478-0895},
	url = {http://www.tandfonline.com/doi/abs/10.1191/1478088706qp063oa},
	doi = {10.1191/1478088706qp063oa},
	language = {en},
	number = {2},
	urldate = {2023-01-15},
	journal = {Qualitative Research in Psychology},
	author = {Braun, Virginia and Clarke, Victoria},
	month = jan,
	year = {2006},
	pages = {77--101},
}

@inproceedings{burns_who_2023,
	address = {Hamburg Germany},
	title = {Who {Do} {We} {Mean} {When} {We} {Talk} {About} {Visualization} {Novices}?},
	isbn = {978-1-4503-9421-5},
	url = {https://dl.acm.org/doi/10.1145/3544548.3581524},
	doi = {10.1145/3544548.3581524},
	language = {en},
	urldate = {2024-02-28},
	booktitle = {Proceedings of the 2023 {CHI} {Conference} on {Human} {Factors} in {Computing} {Systems}},
	publisher = {ACM},
	author = {Burns, Alyxander and Lee, Christiana and Chawla, Ria and Peck, Evan and Mahyar, Narges},
	month = apr,
	year = {2023},
	pages = {1--16},
}

@article{subramanian_towards_2023,
	title = {Towards enhancing ecological validity in user studies: a systematic review of guidelines and implications for {QoE} research},
	volume = {8},
	issn = {2366-0139, 2366-0147},
	shorttitle = {Towards enhancing ecological validity in user studies},
	url = {https://link.springer.com/10.1007/s41233-023-00059-2},
	doi = {10.1007/s41233-023-00059-2},
	language = {en},
	number = {1},
	urldate = {2025-10-31},
	journal = {Quality and User Experience},
	author = {Subramanian, Sruti and De Moor, Katrien and Fiedler, Markus and Koniuch, Kamil and Janowski, Lucjan},
	month = dec,
	year = {2023},
	pages = {6},
}

\end{document}